\documentclass[aps,prb,twocolumn,amsmath,amssymb,showpacs,superscriptaddress]{revtex4}

\usepackage{graphicx}
\usepackage{dcolumn}
\usepackage{bm}

\begin{document}

\title{Surface polaritons on left-handed spheres}

\author{St\'ephane Ancey}
\email{ancey@univ-corse.fr}
\affiliation{ UMR CNRS 6134 SPE, Equipe Ondes et Acoustique, \\
Universit\'e de Corse, Facult\'e des Sciences, Bo{\^\i}te Postale
52, 20250 Corte, France}

\author{Yves D\'ecanini}
\email{decanini@univ-corse.fr}
\affiliation{ UMR CNRS 6134 SPE,
Equipe Physique Semi-Classique (et) de la Mati\`ere Condens\'ee,
\\ Universit\'e de Corse, Facult\'e des Sciences, Bo{\^\i}te
Postale 52, 20250 Corte, France}

\author{Antoine Folacci}
\email{folacci@univ-corse.fr}
\affiliation{ UMR CNRS 6134 SPE,
Equipe Physique Semi-Classique (et) de la Mati\`ere Condens\'ee,
\\ Universit\'e de Corse, Facult\'e des Sciences, Bo{\^\i}te
Postale 52, 20250 Corte, France}

\author{Paul Gabrielli}
\email{gabrieli@univ-corse.fr}
\affiliation{ UMR CNRS 6134 SPE,
Equipe Ondes et Acoustique,
\\ Universit\'e de Corse, Facult\'e des Sciences, Bo{\^\i}te
Postale 52, 20250 Corte, France}

\date{\today}

\begin{abstract}

We consider the interaction of an electromagnetic field with a
left-handed sphere, i.e., with a sphere fabricated from a
left-handed material, in the framework of complex angular momentum
techniques. We emphasize more particularly, from a semiclassical
point of view, the resonant aspects of the problem linked to the
existence of surface polaritons. We prove that the long-lived
resonant modes can be classified into distinct families, each family
being generated by one surface polariton propagating close to the
sphere surface and we physically describe all the surface polaritons
by providing, for each one, its dispersion relation and its damping.
This can be achieved by noting that each surface polariton
corresponds to a particular Regge pole of the electric part (TM) or
the magnetic part (TE) of the $S$ matrix of the sphere. Moreover,
for both polarizations, we find that there exists a particular
surface polariton which corresponds, in the large radius limit, to
that supported by the plane interface. There also exists, for both
polarizations, an infinite family of surface polaritons of
whispering gallery type having no analogs in the plane interface
case and specific to left-handed materials. They present a
``left-handed behavior" (phase and group velocities are opposite) as
well as a very weak damping. They could be very useful in the
context of plasmonics or cavity quantum electrodynamics.

\end{abstract}

\pacs{78.20.Ci, 41.20.Jb, 73.20.Mf, 42.25.Fx}

\maketitle

\section{Introduction}

In the present paper, we continue our analysis of the scattering of
electromagnetic waves by objects of simple shape made of a
left-handed material. (We refer to
Refs.~\onlinecite{Veselago,Smith2000,Smith2000a,Smith2001a,Smith2001b,PendrySmith2004}
for important articles dealing with such materials.) Our analysis is
based on the complex angular momentum (CAM) method which allows us
to completely describe the resonant aspects of the problem. In a
previous work, we have studied scattering from a left-handed
cylinder (disk) \cite{ADFG_2}. In the present paper, we shall extend
this analysis to scattering from a left-handed sphere because of its
numerous potential practical applications.

Scattering from left-handed spheres has been already considered in
some recent
articles\cite{Ruppin2000,Klimov2002,Shen1_2003,Shen2_2003,
RaabeETAL2003,RamakrishnaPendry04,GaoETAL2004,Monzon2004,LiuETAL2004,Vial2006}
and the aspects linked to the existence of the resonant surface
polariton modes (RSPM's) supported by left-handed spheres have been
more particularly considered in
Refs.~\onlinecite{Ruppin2000,Klimov2002,Vial2006}. In the present
paper, by using CAM techniques in connection with asymptotics beyond
all orders\cite{Dingle73,Berry89,BerryHowls90,SegurTL91}, we shall
provide a clear physical explanation for the excitation mechanism of
the RSPM's of the sphere as well as a simple mathematical
description of the surface waves, i.e., of the so-called surface
polaritons (SP's), that generate them. We refer to the Introduction
of Ref.~\onlinecite{ADFG_1} for a presentation of the CAM method and
for a short bibliography as well as to the monographs of
Newton\cite{New82}, Nussenzveig\cite{Nus92} and Grandy\cite{Grandy}
for more details. It should be noted that all the techniques which
we use here are well-known techniques of electromagnetism of
ordinary dielectric media. It is only recently that they have been
introduced in the context of electromagnetism of dispersive media
(see Refs.~\onlinecite{ADFG_1,ADFG_2,ADFG_3}). They enable us to go
well beyond the work completed until now to describe the SP's
propagating on objects of simple shape.

Our paper is organized as follows. Section II is devoted to the
exact theory: we introduce our notations, we provide the expression
of the $S$ matrix of the system and we then discuss the resonant
aspects of the problem for both polarizations. In Sec. III, by using
CAM techniques, we qualitatively describe the SP's supported by the
left-handed sphere and we establish the connection between these
SP's and the associated RSPM's. In Sec. IV, by using asymptotic
techniques, we describe semiclassically the different SP's and we
provide analytic expressions for their dispersion relations and
their damping. We show more particularly the existence of SP's of
whispering gallery type. Finally, in Sec. V, we conclude our paper
by emphasizing the main results of our work and by briefly
discussing the implication of some of our results in the context of
plasmonics and cavity quantum electrodynamics.

\section{Exact $S$ matrices and scattering resonances}

\subsection{General theory}

From now on, we consider the interaction of an electromagnetic field
with a sphere of radius $a$ fabricated from a metamaterial and
having an effective frequency-dependent permittivity $\epsilon
(\omega)$ and an effective frequency-dependent permeability $\mu
(\omega)$. Here, and in the following, we implicitly assume the time
dependence $\exp(-i\omega t)$ for electric and magnetic fields. We
consider that the sphere is embedded in a host medium of infinite
extent having the electromagnetic properties of the vacuum. We
introduce the usual spherical coordinate system $(r ,\theta
,\varphi)$. It is chosen so that the sphere and surrounding medium
respectively occupy the regions corresponding to the range $0 \le r
< a$ (region II) and to the range $r > a$ (region I). Furthermore,
in order to describe wave propagation, we also introduce the wave
number
\begin{equation}
k(\omega)=\frac{\omega }{c},
\end{equation}
where $c$ denotes the velocity of light in vacuum, and the
refractive index of the sphere
\begin{equation}
n(\omega)=\sqrt{ \epsilon (\omega) \mu (\omega)}.
\end{equation}

As far as the electric permittivity $\epsilon (\omega)$ and the
magnetic permeability $\mu (\omega)$ of the sphere are concerned, we
choose the expressions already used in Ref.~\onlinecite{ADFG_2} for
the cylinder. We assume that they are respectively given by
\begin{equation}
\epsilon (\omega) = 1- \frac{\omega_p^2}{\omega ^2} \label{PetP1}
\end{equation}
and
\begin{equation}
\mu (\omega) =  1 - \frac{F \omega^2}{\omega ^2 - \omega_0^2} =(1-F)
\left( \frac{\omega^2 - \omega_b^2 }{\omega ^2 - \omega_0^2}
\right), \label{PetP2}
\end{equation}
where $0<F<1$ and $\omega_b=\omega_0/\sqrt{1-F}$. We are aware that
these expressions do not describe the electromagnetic behavior of
all the real left-handed media but they are the most used in the
literature. Moreover, even if we considered more complicated
expressions for the electric permittivity and magnetic permeability,
our main results would remain valid (see the Conclusion of
Ref.~\onlinecite{ADFG_2}).

For the theoretical aspects of our work, we shall assume that
$\omega_0 < \omega_b < \omega_p$. We have $\epsilon (\omega) <0$ in
the frequency range $\omega \in \left]0, \omega_p \right[$ and $\mu
(\omega) <0$ in the frequency range $\omega \in \left]\omega_0,
\omega_b \right[$. Thus, the electric permittivity, the magnetic
permeability and the refractive index are simultaneously negative in
the region $\omega_0 < \omega < \omega_b$. In that region, the
metamaterial presents a left-handed behavior. As far as the
numerical aspects of our work are concerned, we shall work with
$F=0.4$ and with the reduced frequencies $\omega_0a/c=5.52$,
$\omega_ba/c \approx 7.127$ and $\omega_pa/c=11.04$. Even though we
restrict ourselves to that configuration, the results we shall
obtain numerically are, in fact, very general and they permit us to
correctly illustrate the theory. Furthermore, we have chosen these
particular values, which have been already used in
Ref.~\onlinecite{ADFG_2} for the cylinder, in order to be able to
compare the three-dimensional problem with the two-dimensional one.

The $S$ matrix of the sphere is of fundamental importance because it
contains all the information about the interaction of the sphere
with the electromagnetic field. It can be obtained from Maxwell's
equations and usual continuity conditions for the electric and
magnetic fields at the interface between regions I and II
\cite{Stratton,Nus92,Grandy}. Because of the spherical symmetry of
the scatterer, the $S$ matrix is diagonal and its elements $S_{\ell
\ell'}(\omega)$ are given by $S_{\ell \ell'}(\omega)=S_\ell
(\omega)\ \delta _{\ell \ell'}$. For our problem, the elements of
the electric part (TM polarization) of the $S$ matrix are given by
\begin{equation}
S_\ell^E(\omega)=  1 -2a_\ell^E(\omega) \label{SE1}
\end{equation}
with
\begin{equation}
a_\ell^E(\omega)= \frac{C^E_\ell(\omega)}{D^E_\ell(\omega)},
\label{SE2}
\end{equation}
where $C^E_\ell(\omega)$ and $D^E_\ell(\omega)$ are two $2\times 2$
determinants which are explicitly given by
\begin{subequations}\label{SE3ab}
\begin{eqnarray}
C_{\ell }^{E}(\omega ) &=&\sqrt{\varepsilon \left( \omega \right)
/\mu \left( \omega \right) }\psi _{\ell }\left[ n\left( \omega
\right) \omega a/c\right] \psi _{\ell }^{\prime }\left( \omega
a/c\right) \nonumber \\
& & \quad -\psi _{\ell }\left( \omega a/c\right) \psi _{\ell
}^{\prime }\left[ n\left( \omega
\right) \omega a/c\right]  \label{SE3a} \\
D_{\ell }^{E}(\omega ) &=&\sqrt{\varepsilon \left( \omega \right)
/\mu \left( \omega \right) }\psi _{\ell }\left[ n\left( \omega
\right) \omega a/c\right] \zeta _{\ell }^{(1)^{\prime }}\left(
\omega a/c\right) \nonumber \\
& & \quad -\zeta _{\ell }^{(1)}\left( \omega a/c\right) \psi _{\ell
}^{\prime }\left[ n\left( \omega \right) \omega a/c\right],
\label{SE3b}
\end{eqnarray}%
\end{subequations}
while the elements of its magnetic part (TE polarization) are given
by
\begin{equation}
S_\ell^M(\omega)=  1-2a_\ell^M(\omega) \label{SM1}
\end{equation}
with
\begin{equation}
a_\ell^M(\omega)= \frac{C^M_\ell(\omega)}{D^M_\ell(\omega)},
\label{SM2}
\end{equation}
where $C^M_\ell(\omega)$ and $D^M_\ell(\omega)$ are also two
$2\times 2$ determinants which are explicitly given by
\begin{subequations}\label{SM3ab}
\begin{eqnarray}
C_{\ell }^{M}(\omega ) &=&\sqrt{\varepsilon \left( \omega \right)
/\mu \left( \omega \right) }\psi _{\ell }\left( \omega a/c\right)
\psi _{\ell }^{\prime }\left[ n\left( \omega \right) \omega
a/c\right] \nonumber \\
& & \quad -\psi _{\ell }\left[ n\left( \omega \right) \omega
a/c\right] \psi _{\ell }^{\prime
}\left( \omega a/c\right)  \label{SM3a}\\
D_{\ell }^{M}(\omega ) &=&\sqrt{\varepsilon \left( \omega \right)
/\mu \left( \omega \right) }\zeta _{\ell }^{(1)}\left( \omega
a/c\right) \psi _{\ell }^{\prime }\left[ n\left( \omega \right)
\omega a/c\right] \nonumber \\
& & \quad -\psi _{\ell }\left[ n\left( \omega \right) \omega
a/c\right] \zeta _{\ell }^{(1)^{\prime }}\left( \omega a/c\right).
\label{SM3b}
\end{eqnarray}
\end{subequations}
In Eqs.~(\ref{SE3ab}) and (\ref{SM3ab}), we use the Ricatti-Bessel
functions $\psi_\ell(z)$ and $\zeta _{\ell }^{(1)}(z)$ which are
linked to the spherical Bessel functions $j_\ell(z)$ and
$h^{(1)}_\ell(z)$ by $\psi_\ell(z)=zj_\ell(z)$ and $\zeta _{\ell
}^{(1)}(z)=zh^{(1)}_\ell(z)$ (see Ref.~\onlinecite{AS65}).

From the $S$ matrix elements, we can, in particular, construct the
total scattering cross section of the sphere which is defined as the
ratio of the total energy scattered and the incident energy
intercepted. It can be expressed in terms of the coefficients
$a_\ell^E(\omega)$ and $a_\ell^M(\omega)$ and it is given
by\cite{Stratton,Nus92,Grandy}
\begin{equation}\label{crosssection}
\sigma_T(\omega) =  \frac{2\pi}{\left[k \left( \omega
\right)\right]^2}\sum_{\ell=1}^{\infty}
(2\ell+1)[|a_\ell^E(\omega)|^2+|a_\ell^M(\omega)|^2].
\end{equation}

From the $S$ matrix elements, we can also precisely describe the
resonant behavior of the sphere as well as the geometrical and
diffractive aspects of the scattering process. Here, the dual
structure of the $S$ matrix plays a crucial role. Indeed, the $S$
matrix is a function of both the frequency $\omega$ and the angular
momentum index $\ell$. It can be analytically extended into the
complex $\omega$ plane as well as into the complex $\lambda$ plane
(the CAM plane). (Here, $\lambda$ denotes the complex angular
momentum index replacing $\ell +1/2$ where $\ell$ is the ordinary
momentum index\cite{New82,Nus92,Grandy}.) The poles of the $S$
matrix lying in the fourth quadrant of the complex $\omega$ plane
are the complex frequencies of the resonant modes. These resonances
are determined by solving
\begin{equation}\label{det}
D^{E,M}_\ell(\omega)=0 \quad \mathrm{for} \quad \ell \in
\mathbf{N}^*.
\end{equation}
The solutions of (\ref{det}) are denoted by $\omega_{\ell
p}=\omega^{(o)}_{\ell p}-i\Gamma _{\ell p}/2$ where
$\omega^{(o)}_{\ell p}>0$ and $\Gamma _{\ell p}>0$, the index $p$
permitting us to distinguish between the different roots of
(\ref{det}) for a given $\ell$. In the immediate neighborhood of the
resonance $\omega_{\ell p}$, $S^{E,M}_\ell(\omega)$ has the
Breit-Wigner form, i.e., proportional to
\begin{equation}\label{BW}
\frac{\Gamma _{\ell p}/2}{\omega -\omega^{(o)}_{\ell p}+i\Gamma
_{\ell p}/2}.
\end{equation}
The structure of the $S$ matrix in the complex $\lambda$ plane
allows us, by using integration contour deformations, Cauchy's
theorem and asymptotic analysis, to provide a semiclassical
description of scattering in terms of rays (geometrical and
diffracted). In that context, the poles of the $S$-matrix lying in
the CAM plane (the so-called Regge poles) are associated with
diffraction. They are determined by solving
\begin{equation}\label{RP}
D^{E,M}_{\lambda-1/2}( \omega)=0 \quad \mathrm{for} \quad \omega >
0.
\end{equation}
Of course, when a connection between these two faces of the
$S$-matrix can be established, the resonance aspects of scattering
are then semiclassically interpreted.

\begin{figure}
\includegraphics[height=7cm,width=8.6cm]{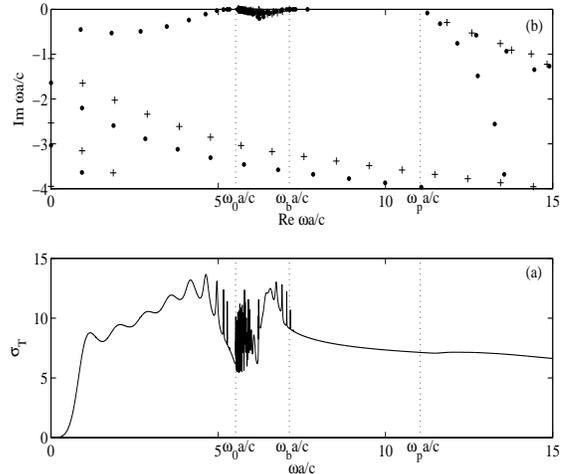}
\caption{\label{fig:general_TETM} a) Total cross section $\sigma_T$.
b) Scattering resonances in the complex $\omega a/c$ plane. Dots
$(\cdot)$ correspond to poles of $S^E(\omega)$ while plus ($+$)
correspond to poles of $S^M(\omega)$.}
\end{figure}

\subsection{Numerical results and physical interpretation}

In Fig.~\ref{fig:general_TETM}a, we display the total cross section
plotted as a function of the reduced frequency $\omega a /c$. Rapid
variations of sharp characteristic shapes can be observed. Such a
strongly fluctuating behavior is due to the scattering resonances
associated with the long-lived resonant modes of the sphere, i.e.,
the long-lived resonant states of the photon-sphere system: when a
pole of the $S$ matrix is sufficiently close to the real axis in the
complex $\omega$ plane, it has a strong influence on the total cross
section. In Fig.~\ref{fig:general_TETM}b, resonances are exhibited
for the two polarizations. For certain frequencies, we can clearly
observe a one-to-one correspondence between the peaks of $\sigma
_T(\omega)$ and the resonances near the real $\omega a/c$ axis but,
in general, the situation seems very confused. This is due to the
profusion of long-lived resonant modes in and around the frequency
range where the sphere presents a left-handed behavior. In spite of
that, it should be nevertheless noted that (i) the strongly
fluctuating behavior generated by electric resonances (TM
polarization) is localized within and slightly around the frequency
range where the sphere presents a left-handed behavior and (ii) the
strongly fluctuating behavior generated by magnetic resonances (TE
polarization) is totally localized within that frequency range.
Furthermore, by zooming in on the distribution of resonances in
regions close to the real axis of the complex $\omega$ plane, we
have also observed accumulations of resonances for large values of
$\ell$:

\qquad (i) For the TM polarization, there exists an accumulation of
resonances which converges to the limiting frequency $\omega_s$
satisfying
\begin{equation}\label{TMaccSPinf1}
\epsilon (\omega_s) + 1 =0
\end{equation}
and given by
\begin{equation}\label{TMaccSPinf2}
\omega_s =  \frac{\omega_p}{ \sqrt{2}}.
\end{equation}
We have for the corresponding numerical reduced frequency
$\omega_sa/c \approx 7.806$.

\qquad (ii) For the TE polarization, there exists an accumulation of
resonances at the limiting frequency $\omega_f$ satisfying
\begin{equation}\label{TEaccSPinf1}
\mu (\omega_f) + 1 =0
\end{equation}
and given by
\begin{equation}\label{TEaccSPinf2}
\omega_f = \omega_0 \sqrt{\frac{2}{2-F}}.
\end{equation}
We have for the corresponding numerical reduced frequency
$\omega_fa/c \approx 6.172$.

\qquad (iii) For both polarizations, there exists an accumulation of
resonances at the pole $\omega_0$ of $\mu (\omega)$ which
corresponds more precisely to
\begin{equation}\label{TMTEaccWGSP}
\mu (\omega_0) = - \infty.
\end{equation}

From now on, we shall more particularly focus our attention on the
physical interpretation of the long-lived resonant modes whose
excitation frequencies belong to frequency ranges in which $n
(\omega) <0$. We shall prove that the corresponding resonances are
generated by exponentially small attenuated SP's propagating close
to the sphere surface and we shall provide a numerical and a
theoretical description of these surface waves. The corresponding
resonant modes are the so-called RSPM's. For the other resonant
modes associated with bulk polaritons (the resonant modes whose
excitation frequencies belong to the frequency range in which $n
(\omega) > 0$ and those with shorter lifetime), we do not provide a
similar analysis. This is not very serious as they do not have, in
physics, the importance of RSPM's. Indeed, in the new field of
plasmonics, those are the SP's with long propagation lengths and
therefore very small attenuations that are especially interesting
from the point of view of the practical applications. Furthermore,
if we consider the photon-sphere system as an artificial atom for
which the photon plays the usual role of the electron, we must then
keep in mind that, in the scattering of a photon with frequency
$\omega^{(o)}_{\ell p}$, a decaying state (i.e., a quasibound state)
of the photon-sphere system is formed. It has a finite lifetime
proportional to $1/\Gamma _{\ell p}$. The resonant states whose
complex frequencies belong to the family generated by SP's are
therefore the most interesting because they are very long-lived
states.

We shall finally conclude this section by making a brief comparison
with the results obtained in our previous studies concerning the
left-handed cylinder\cite{ADFG_2} and the metallic and
semiconducting spheres\cite{ADFG_3}:

\qquad (i) The total cross sections for the left-handed cylinder and
the left-handed sphere as well as their spectra of resonances are
rather similar. However, it should be noted that in the scattering
by a sphere both the TM and TE polarizations contribute to the total
cross section (we recall that for the cylinder, the two
polarizations can be studied separately). But, it is also important
to note that, for both scatterers, SP's and their associated RSPM's
correspond to only one polarization (the TE polarization for the
cylinder and the TM polarization for the sphere).

\qquad (ii) As far as the spectrum of resonances is concerned, the
left-handed sphere is a physical system much richer than the
metallic or the semiconducting sphere (see Figs. 1 and 2 of
Ref.~\onlinecite{ADFG_3} and the discussion at the end of Sec. II of
that reference) and this is certainly very interesting for practical
applications  of left-handed electromagnetism and more particularly
in plasmonics and in cavity quantum electrodynamics.

\section{Semiclassical analysis: Regge poles, surface polaritons, and resonances}

\subsection{Complex angular momentum approach: Results}

In the CAM approach\cite{New82,Nus92,Grandy}, Regge poles determined
by solving Eq.~(\ref{RP}) are crucial to describe diffraction as
well as resonance phenomena in terms of surface waves. From the
Regge trajectories associated with the SP's supported by the
left-handed sphere, i.e., from the curves $\lambda_\mathrm{SP}
=\lambda_\mathrm{SP} (\omega)$ traced out in the CAM plane by the
corresponding Regge poles as a function of the frequency, we can
more particularly deduce the following:

\quad (i) the dispersion relations of SP's which connect their wave
numbers $k_\mathrm{SP}$ with the frequency $\omega$:
\begin{equation}\label{WNSP}
k_\mathrm{SP} (\omega) = \frac{\mathrm{Re} \, \lambda_\mathrm{SP}
(\omega)}{a},
\end{equation}

\quad (ii) the dampings $\mathrm{Im} \, \lambda_\mathrm{SP}
(\omega)$ of SP's,

\quad (iii) the phase velocities $v_p$ as well as the group
velocities $v_g$ of SP's:
\begin{equation}\label{VpGg}
v_p = \frac{a \omega}{\mathrm{Re} \, \lambda_\mathrm{SP} (\omega)}
\quad \mathrm{and} \quad v_g = \frac{d~a \omega}{d~\mathrm{Re} \,
\lambda_\mathrm{SP} (\omega)},
\end{equation}

 \quad (iv) the semiclassical formula (a Bohr-Sommerfeld-type quantization condition)
 which provides the location of the excitation frequencies
$\omega^{(0)}_{\ell \mathrm{SP}}$ of the resonances generated by
SP's:
\begin{equation}\label{sc1}
\mathrm{Re}  \, \lambda_\mathrm{SP} \left(\omega^{(0)}_{\ell
\mathrm{SP}} \right)= \ell + 1/2,  \qquad \ell =1,2,\dots,
\end{equation}

\quad (v) the semiclassical formula which provides the widths of
these resonances
\begin{equation}\label{sc2} \frac{\Gamma _{\ell
\mathrm{SP}}}{2}= \left.  \frac{\mathrm{Im} \, \lambda_\mathrm{SP}
(\omega )[d/d\omega \, \mathrm{Re}\, \lambda_\mathrm{SP} (\omega )
]}{[d/d\omega \, \mathrm{Re} \, \lambda_\mathrm{SP} (\omega ) ]^2 +
[d/d\omega \, \mathrm{Im} \, \lambda_\mathrm{SP} (\omega ) ]^2 }
\right|_{\omega =\omega^{(0)}_{\ell \mathrm{SP}}}
\end{equation}
and which reduces, in the frequency range where the condition $|
d/d\omega \,\mathrm{Re} \, \lambda_\mathrm{SP} (\omega )  | \gg
|d/d\omega \, \mathrm{Im}\, \lambda_\mathrm{SP} (\omega )  |$ is
satisfied, to
\begin{equation}\label{sc4}
\frac{\Gamma _{\ell \mathrm{SP}}}{2}= \left.  \frac{\mathrm{Im} \,
\lambda_\mathrm{SP} (\omega )}{d/d\omega \, \ \mathrm{Re} \,
\lambda_\mathrm{SP} (\omega )  } \right|_{\omega =\omega^{(0)}_{\ell
\mathrm{SP}}}.
\end{equation}

\noindent All these results can be established by generalizing, {\it
mutatis mutandis}, our approach and our calculations developed in
Refs.~\onlinecite{ADFG_1,ADFG_2} for dispersive cylinders. The
transition from two dimensions to three dimensions induces some
additional technical difficulties (vectorial treatment, existence of
a caustic, asymptotics for spherical harmonics, etc.) which can be
overcome following and extending the works of Newton in quantum
mechanics (see Chap.~13 of Ref.~\onlinecite{New82}) and the works of
Nussenzveig\cite{Nus92} and Grandy\cite{Grandy} in electromagnetism
of ordinary dielectric media.

\begin{figure}
\includegraphics[height=5.4cm,width=8.6cm]{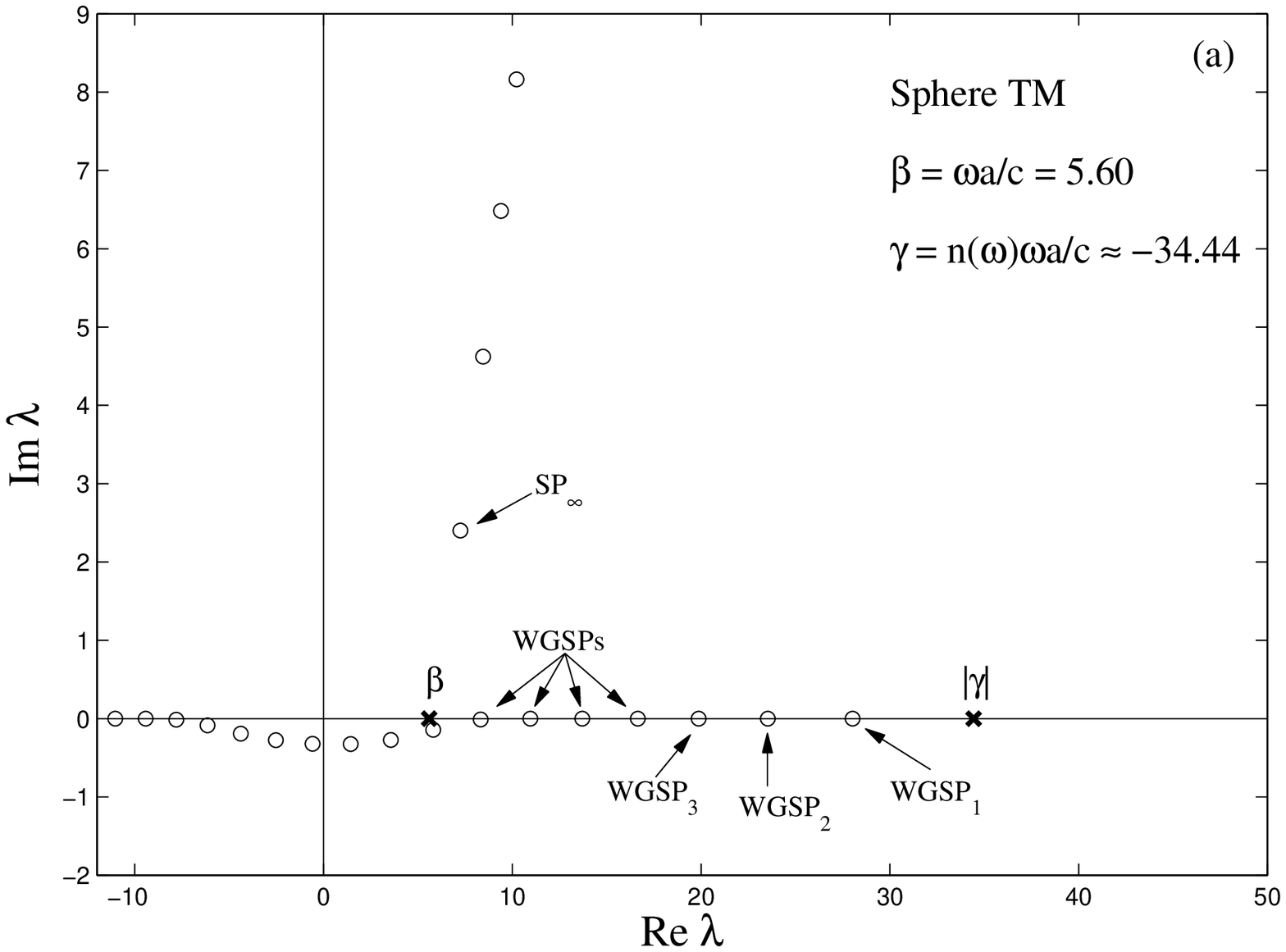}
\includegraphics[height=5.4cm,width=8.6cm]{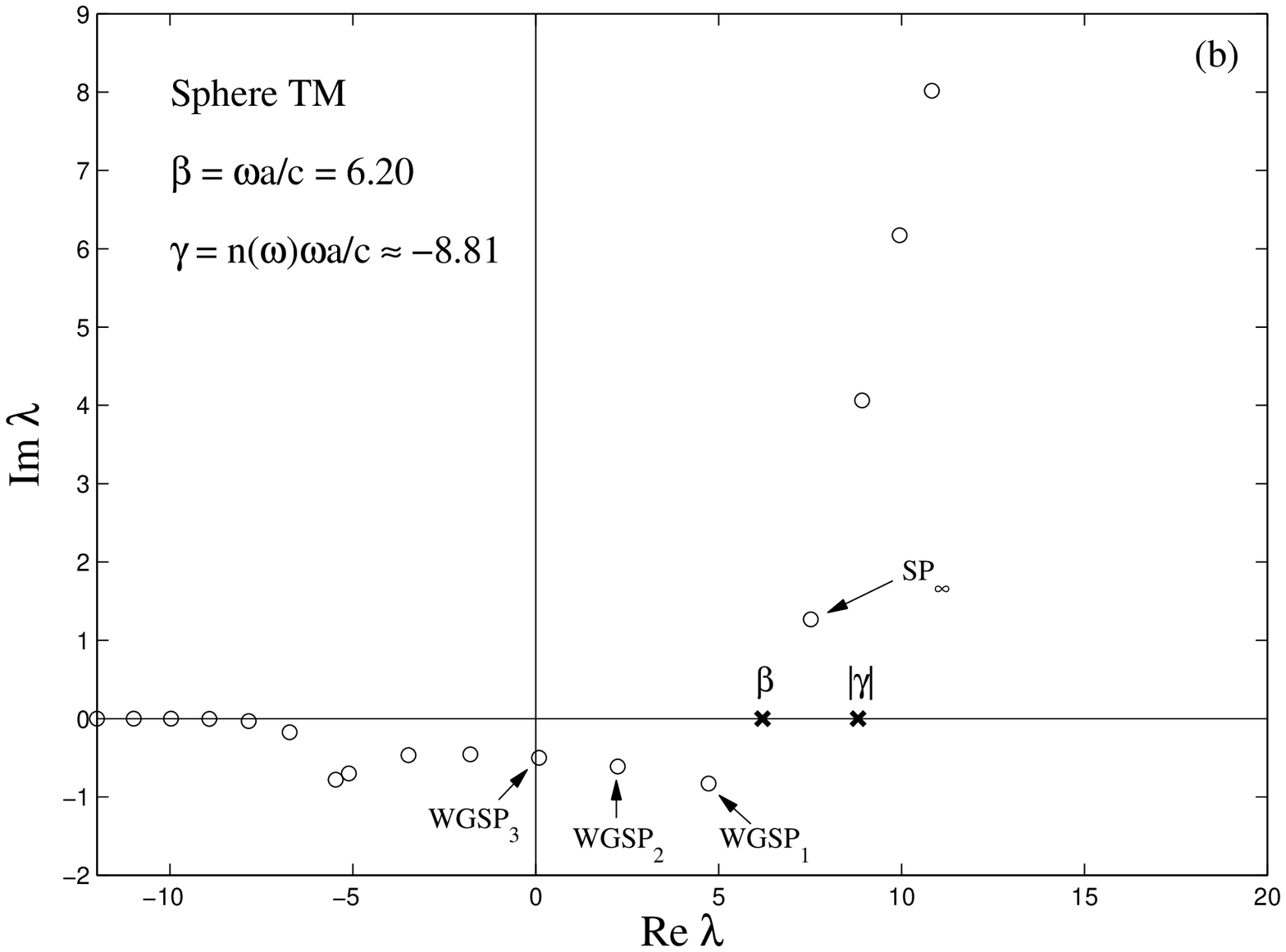}
\includegraphics[height=5.4cm,width=8.6cm]{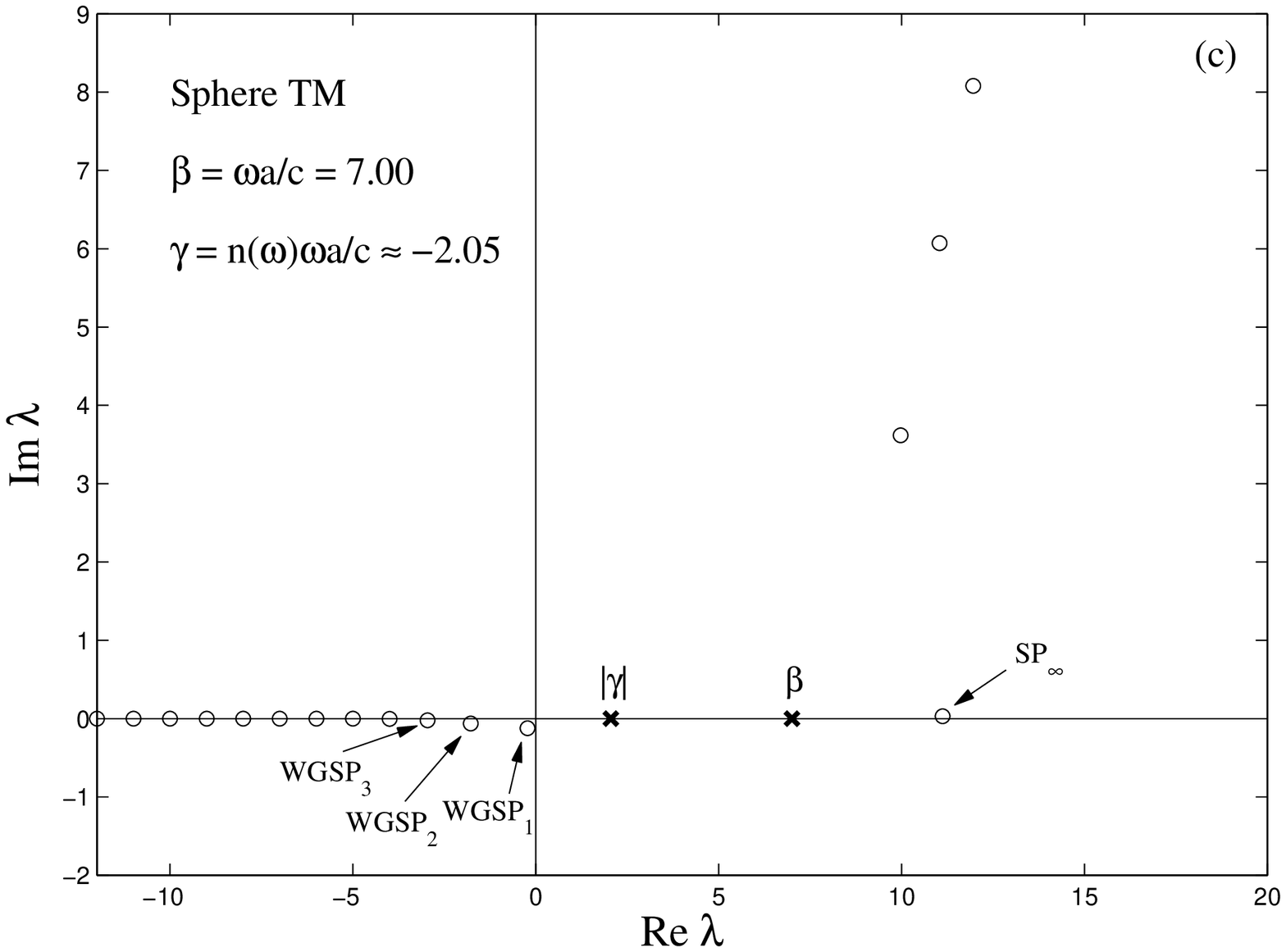}
\caption{\label{fig:RP1} Regge poles in the CAM plane for the TM
polarization. a) The distribution corresponds to $\omega a/c= 5.6$.
b) The distribution corresponds to $\omega a/c= 6.2$. c) The
distribution corresponds to $\omega a/c= 7.0$.}
\end{figure}
\begin{figure}
\includegraphics[height=5.4cm,width=8.6cm]{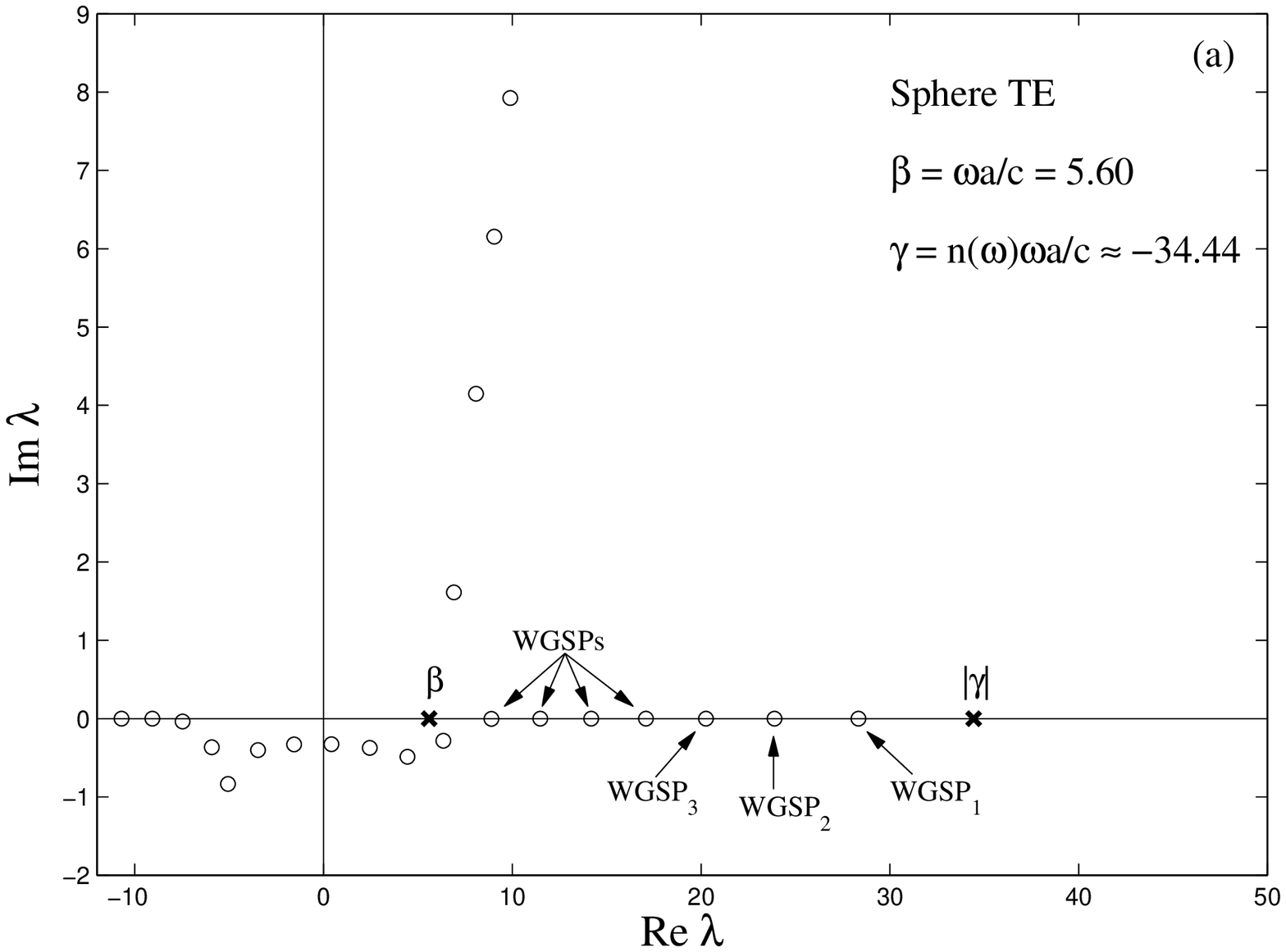}
\includegraphics[height=5.4cm,width=8.6cm]{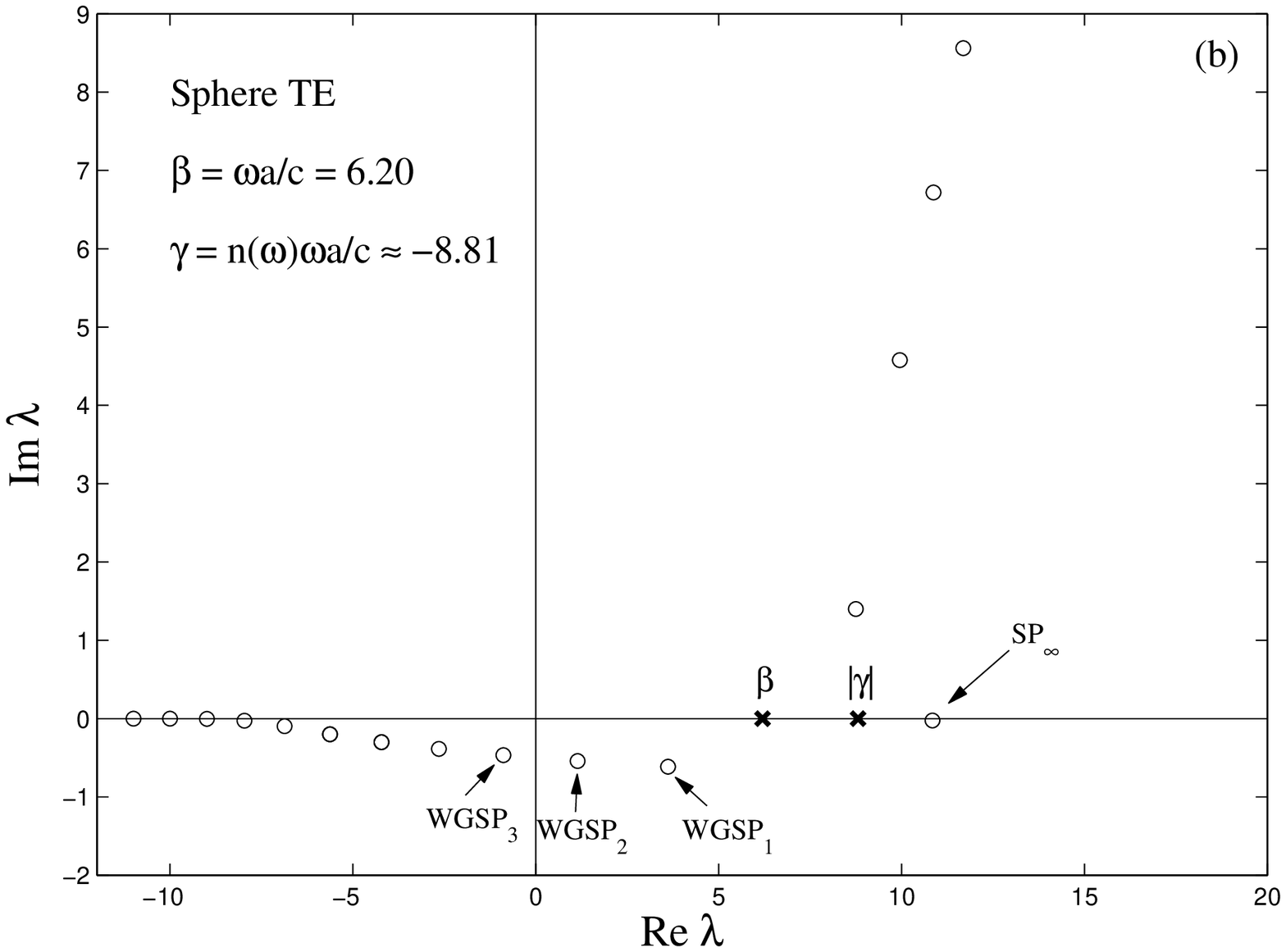}
\includegraphics[height=5.4cm,width=8.6cm]{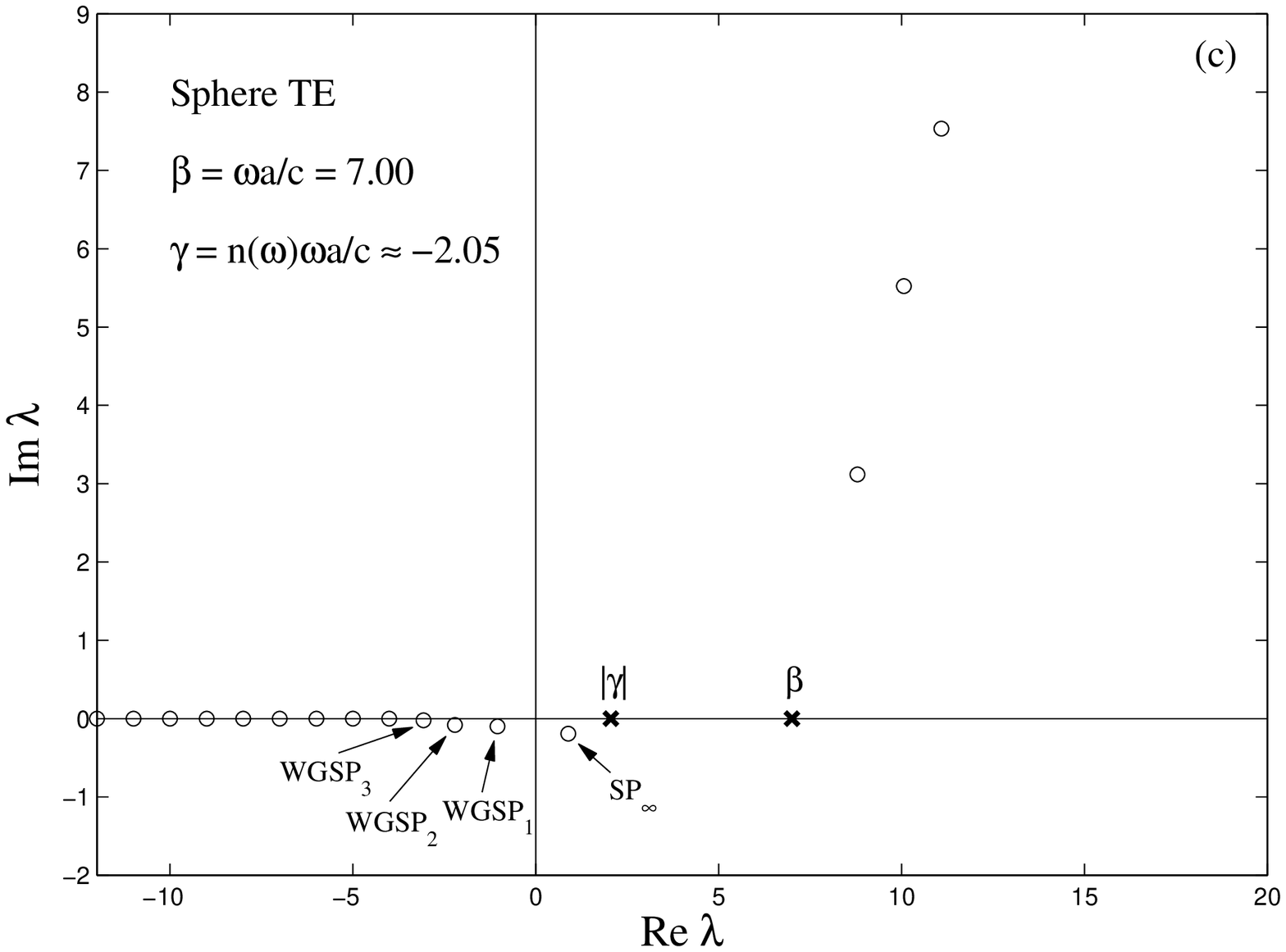}
\caption{\label{fig:RP2} Regge poles in the CAM plane for the TE
polarization. a) The distribution corresponds to $\omega a/c= 5.6$.
b) The distribution corresponds to $\omega a/c= 6.2$. c) The
distribution corresponds to $\omega a/c= 7.0$.}
\end{figure}

\subsection{Regge poles and Regge trajectories}

Figures \ref{fig:RP1} and \ref{fig:RP2} exhibit the distribution of
Regge poles for both polarizations for three different reduced
frequencies lying in the frequency region where $n(\omega) <0$. We
have identified and indicated some particular Regge poles which are
associated with surface waves orbiting around the left-handed sphere
and which explain its resonant behavior. These figures are at first
sight qualitatively similar to Figs.~(4) and (5) of
Ref.~\onlinecite{ADFG_2} where we displayed the corresponding Regge
pole structures for the left-handed cylinder. It should be however
noted that, in the transition from the cylinder to the sphere, the
TM and TE polarizations are exchanged and curvature corrections
induce modifications in the quantitative behavior (as we shall show
in Sec. IV). More precisely, for both polarizations, we can observe
in Figs.~\ref{fig:RP1} and \ref{fig:RP2} the following.

\qquad -- One of these Regge poles is associated with the SP noted
${\mathrm{SP}_\infty}$ which, as we shall show in Sec. IV,
corresponds in the large radius limit (i.e., for $a \to \infty$) to
a SP supported by the plane interface and theoretically described in
Refs.~\onlinecite{RuppinPLA00,Darmanyanetal03,ShadrivovEtAl04}.

\qquad -- The other Regge poles are associated with an infinite
family of SP's of whispering gallery type denoted by
${\mathrm{WGSP}_n}$ with $n \in \mathbb{N}^*$ which have no analogs
for the plane interface (see Sec. IV) as well as for curved metallic
or semiconducting interfaces (see Refs.~\onlinecite{ADFG_1,ADFG_3}).

\begin{figure}
\includegraphics[height=8cm,width=8.6cm]{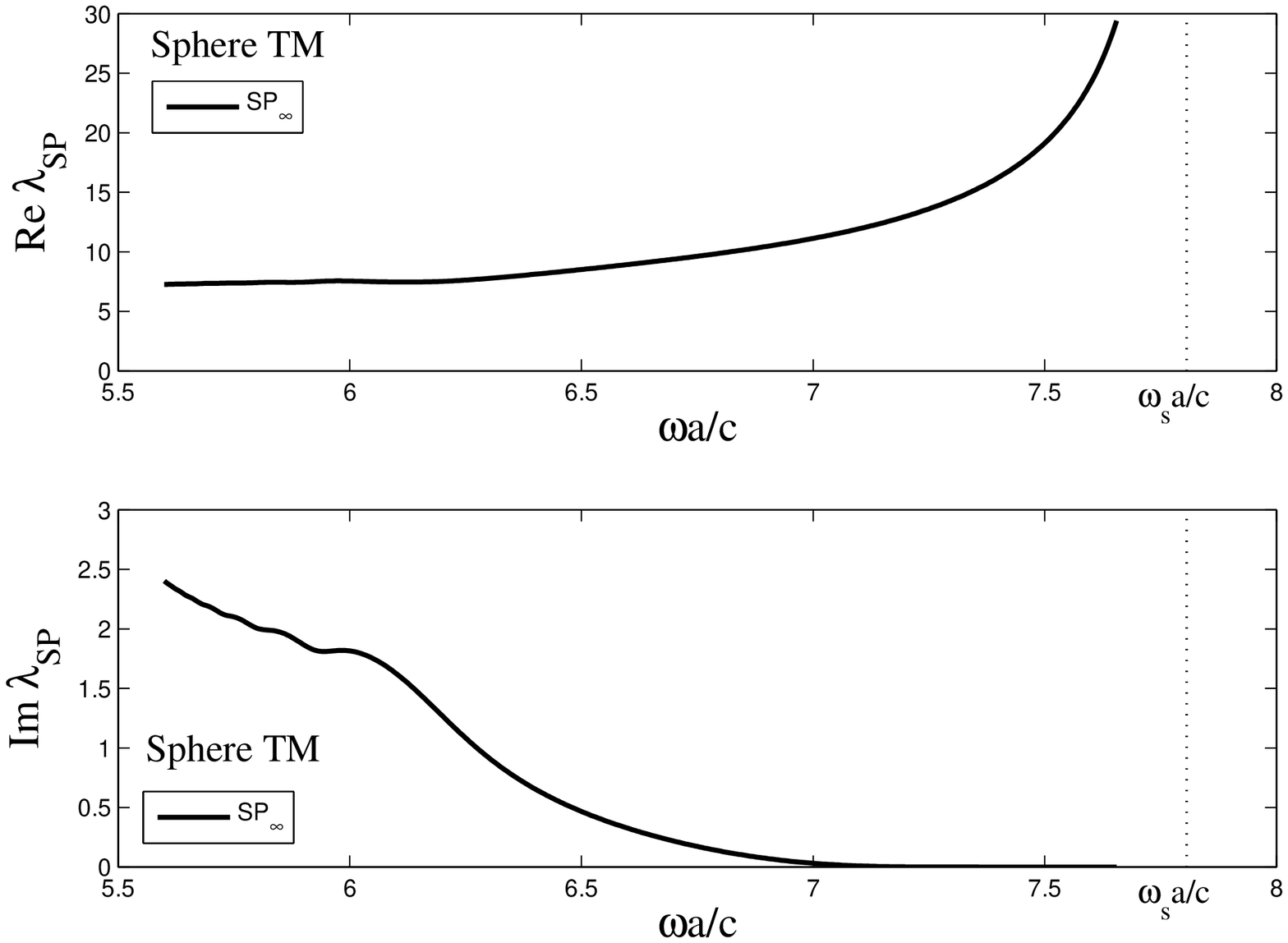}
\caption{\label{fig:RTSPiTM} Regge trajectory for the Regge pole
associated with ${\mathrm{SP}_\infty}$ (TM polarization). }
\includegraphics[height=8cm,width=8.6cm]{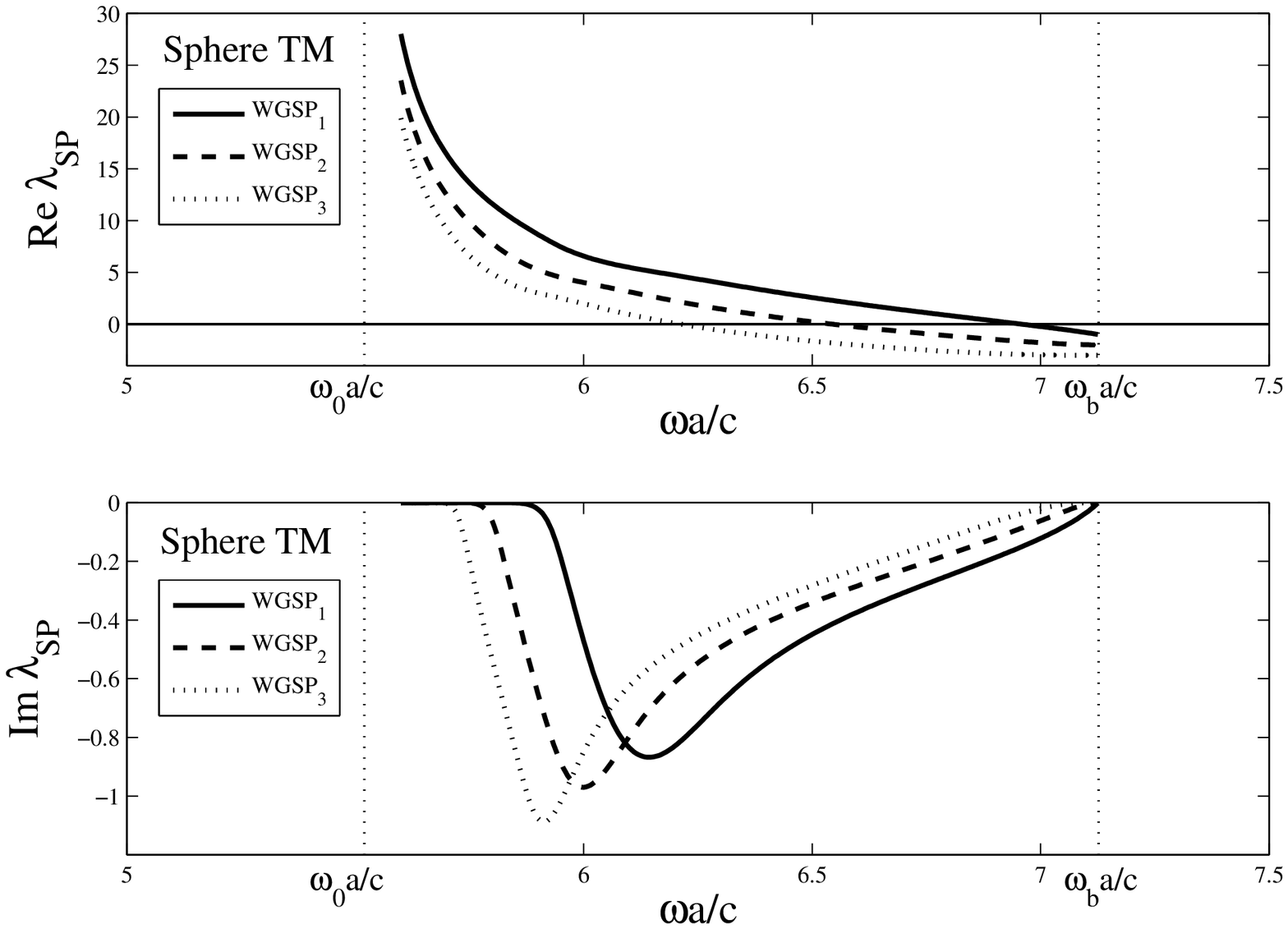}
\caption{\label{fig:RTWGSPTM} Regge trajectories for the Regge poles
associated with the first three whispering-gallery SP's (TM
polarization). }
\end{figure}

\begin{figure}
\includegraphics[height=8cm,width=8.6cm]{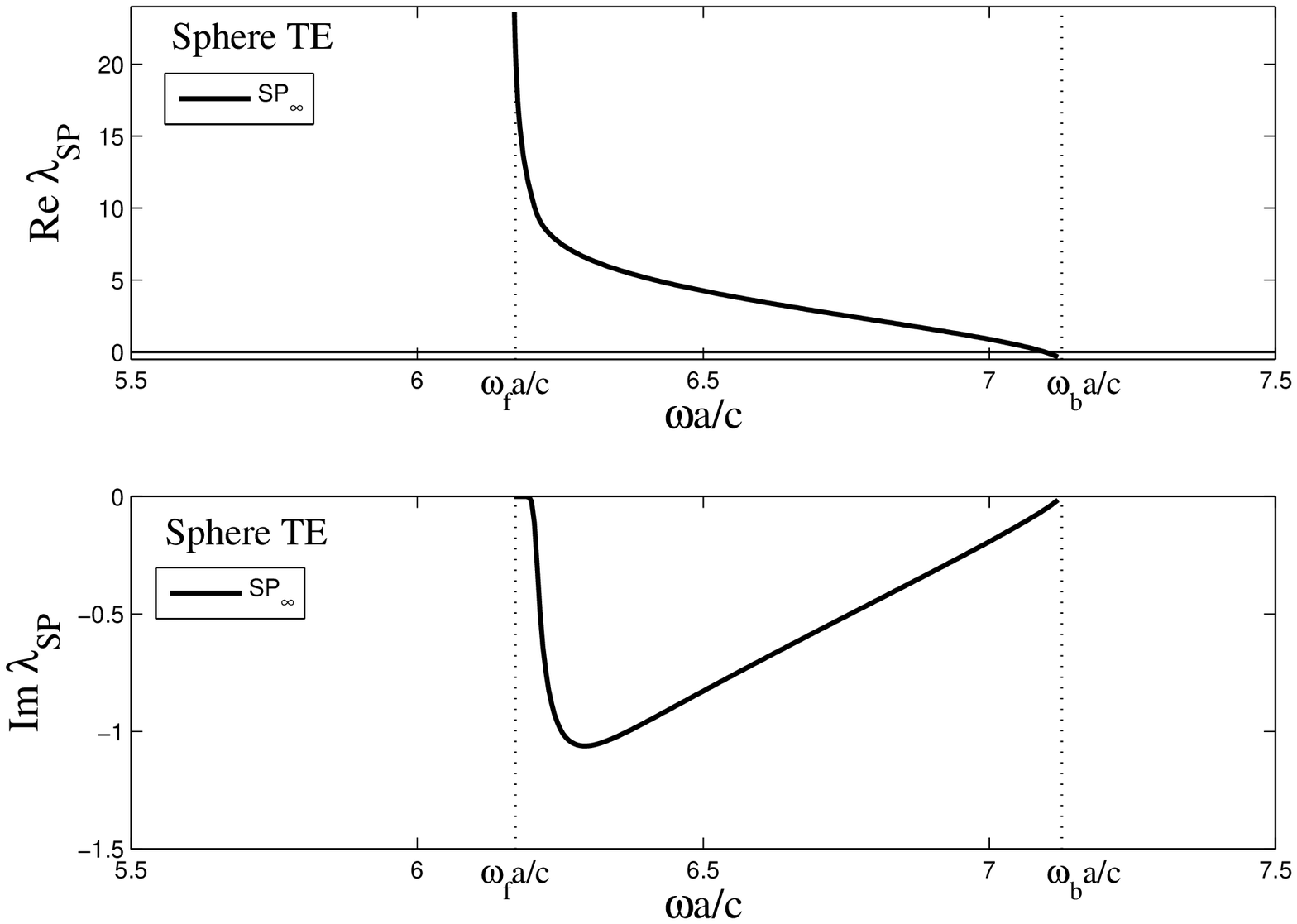}
\caption{\label{fig:RTSPiTE} Regge trajectory for the Regge pole
associated with ${\mathrm{SP}_\infty}$ (TE polarization). }
\includegraphics[height=8cm,width=8.6cm]{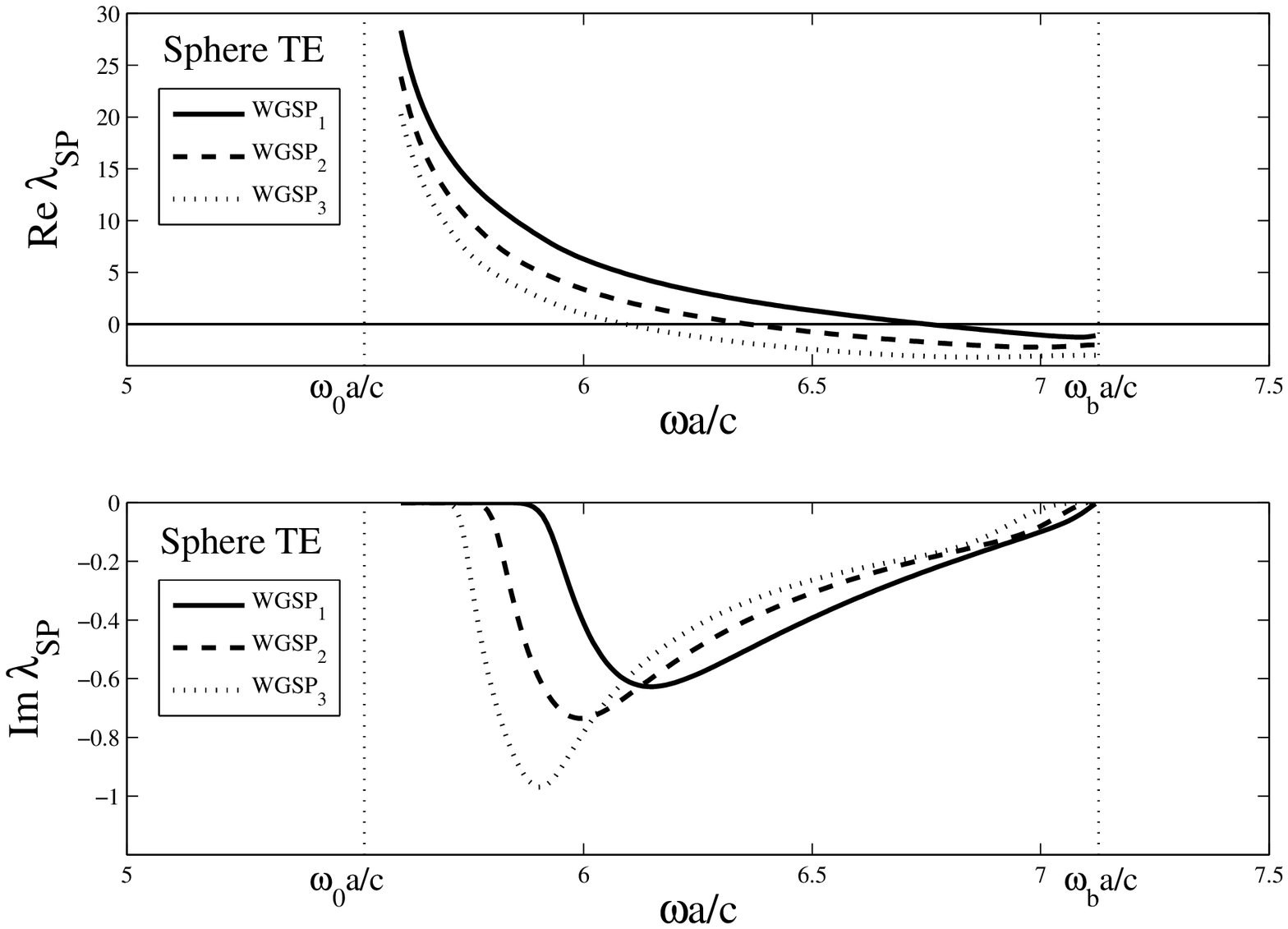}
\caption{\label{fig:RTWGSPTE} Regge trajectories for the Regge poles
associated with the first three whispering-gallery SP's (TE
polarization). }
\end{figure}

\begin{figure}
\includegraphics[height=5.3cm,width=8.6cm]{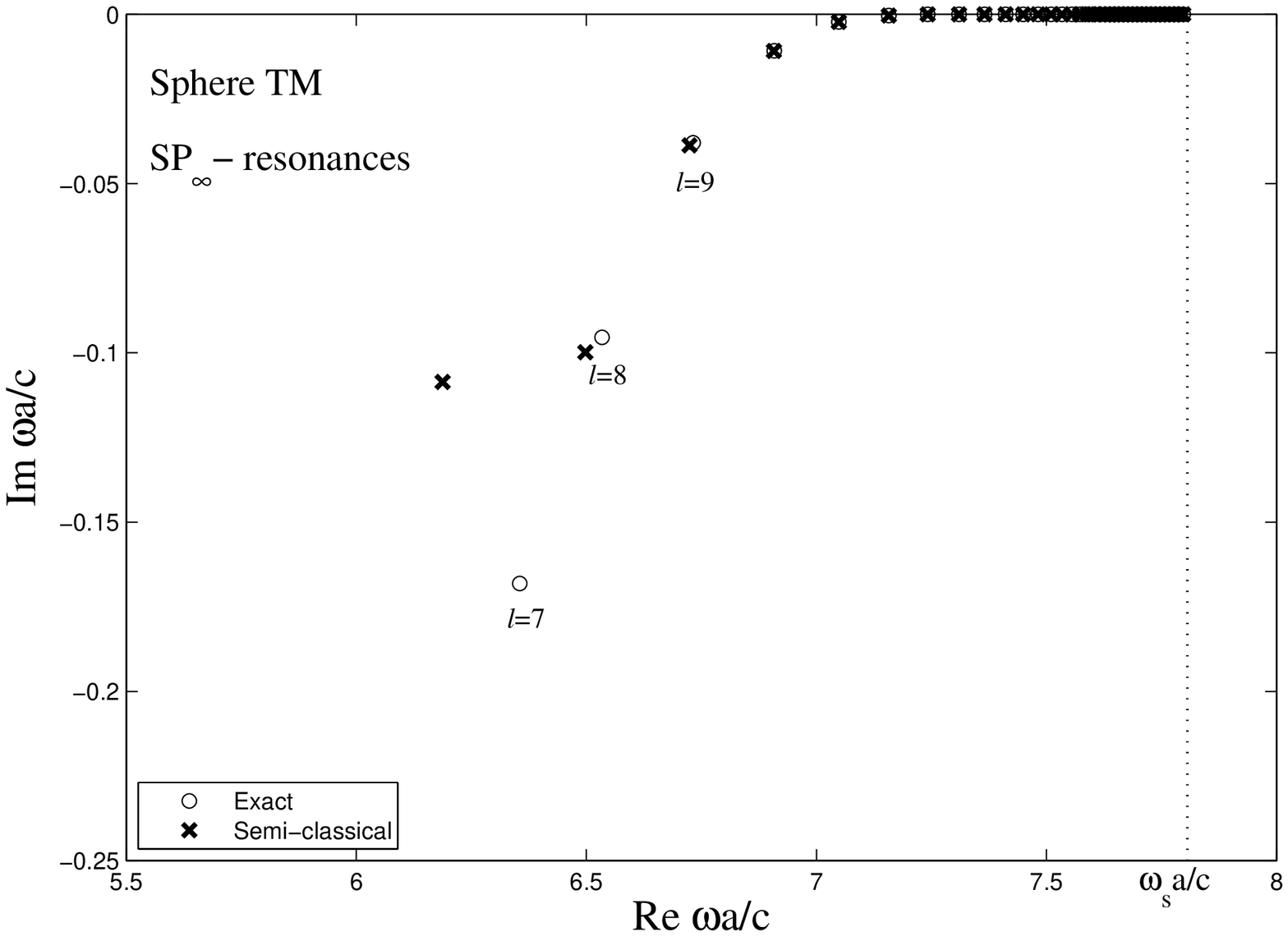}
\caption{\label{fig:ResSPinfTM}  Resonances generated by
${\mathrm{SP}_\infty}$ (TM polarization). }
\includegraphics[height=6.3cm,width=8.6cm]{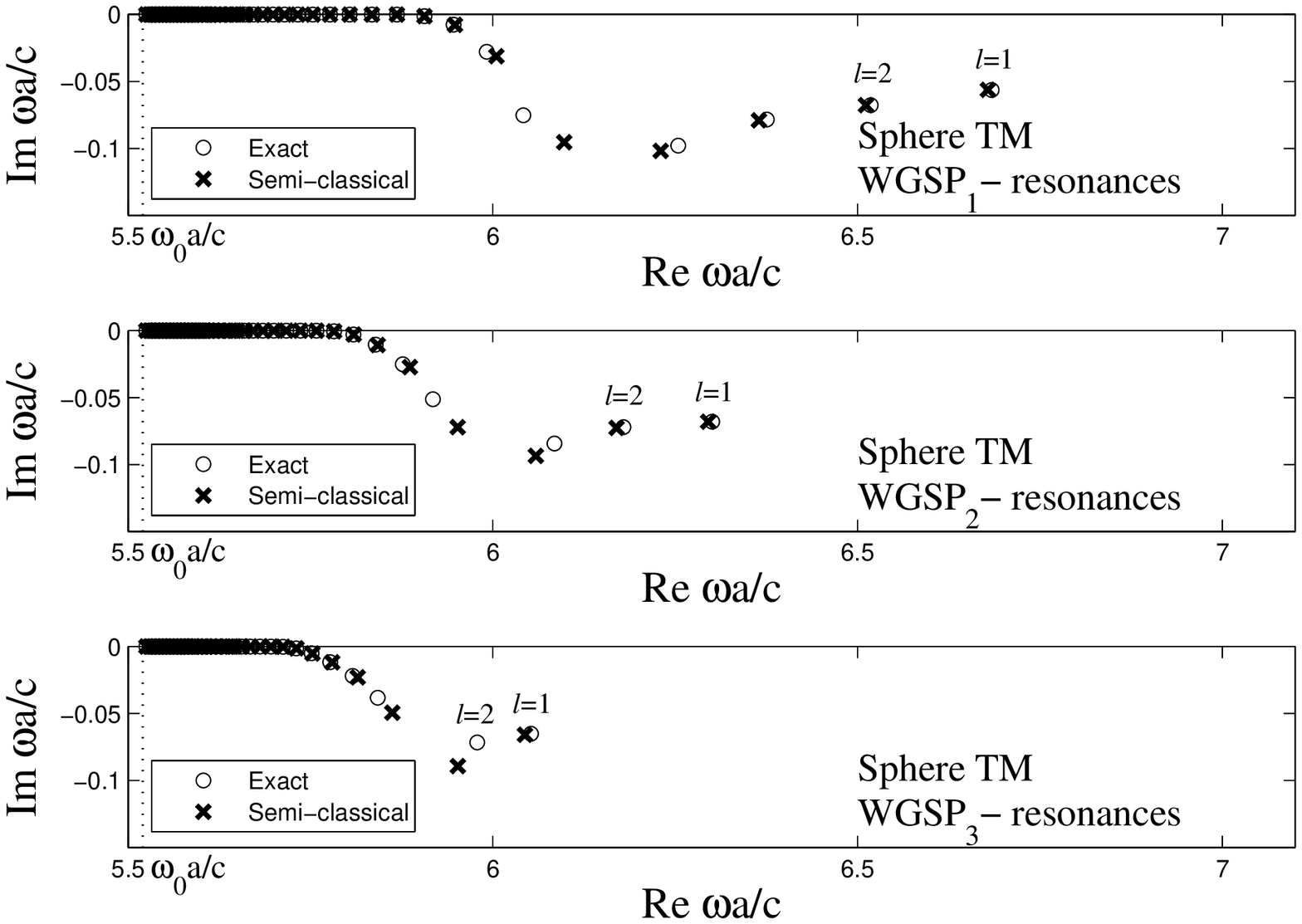}
\caption{\label{fig:ResWGSPTM} Resonances generated by the first
three whispering-gallery SP's (TM polarization). }
\end{figure}
\begin{figure}
\includegraphics[height=5.3cm,width=8.6cm]{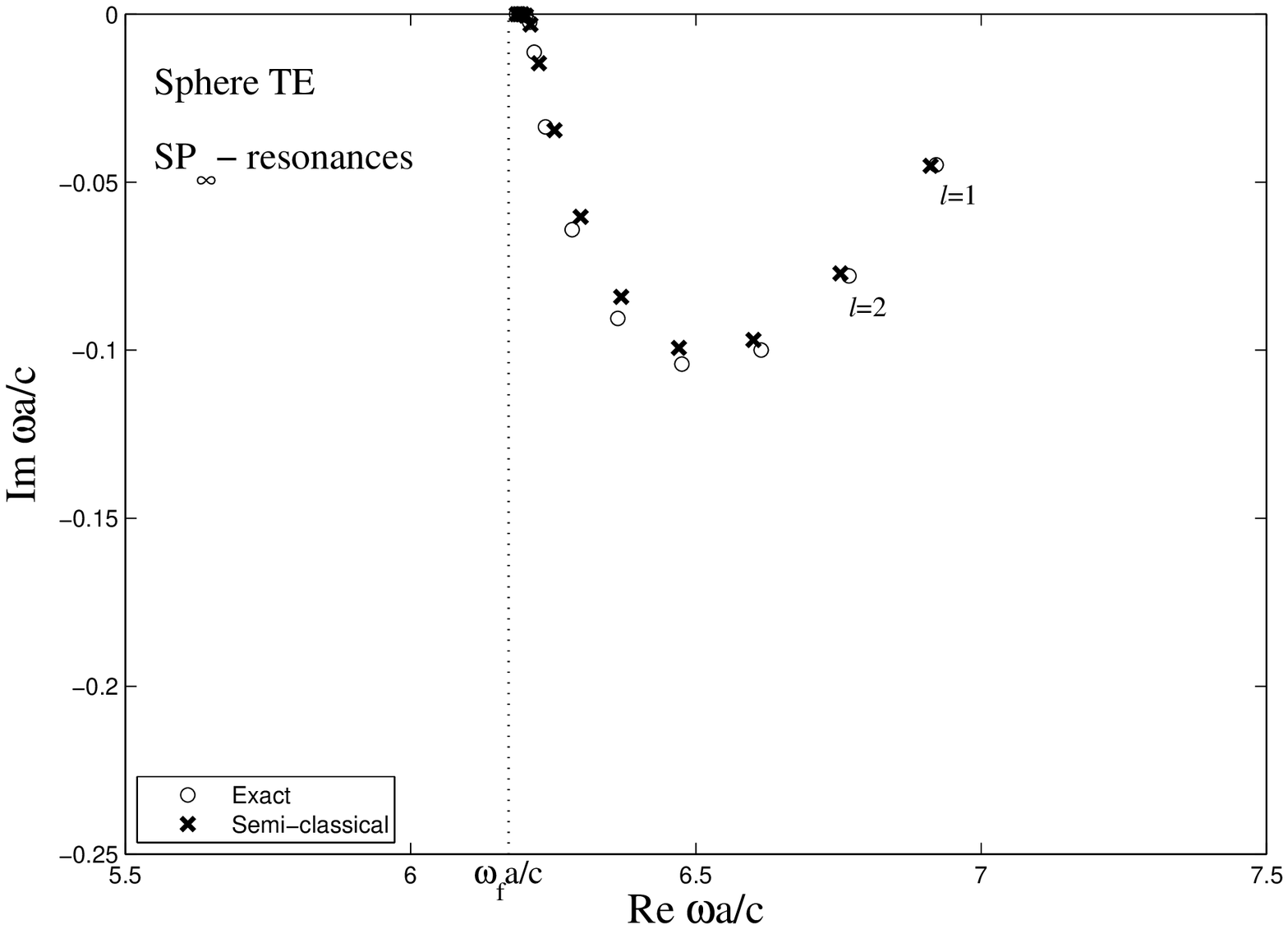}
\caption{\label{fig:ResSPinfTE} Resonances generated by
${\mathrm{SP}_\infty}$ (TE polarization). }
\includegraphics[height=6.3cm,width=8.6cm]{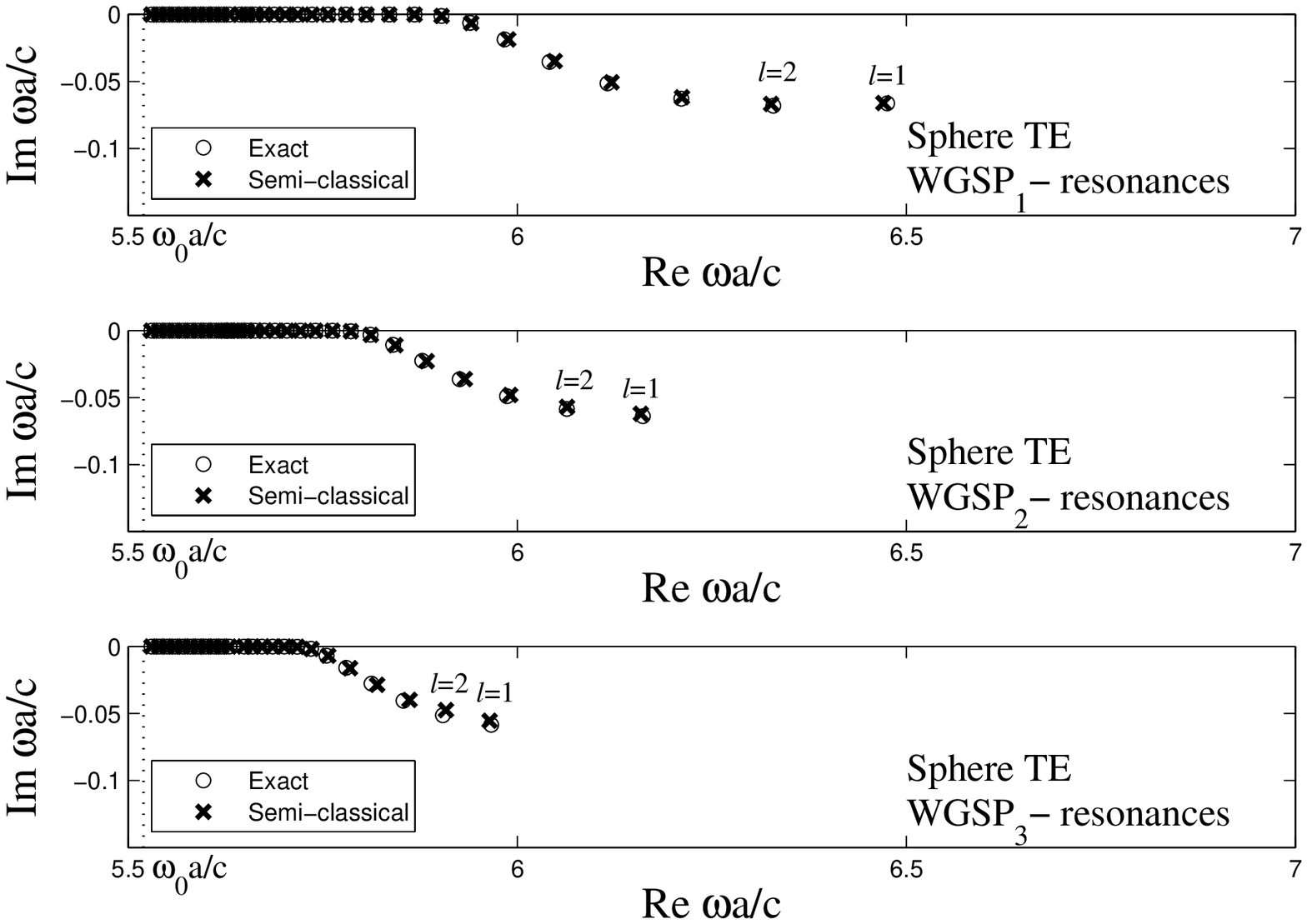}
\caption{\label{fig:ResWGSPTE} Resonances generated by the first
three whispering-gallery SP's (TE polarization). }
\end{figure}
\begin{figure*}
\includegraphics[height=6cm,width=15cm]{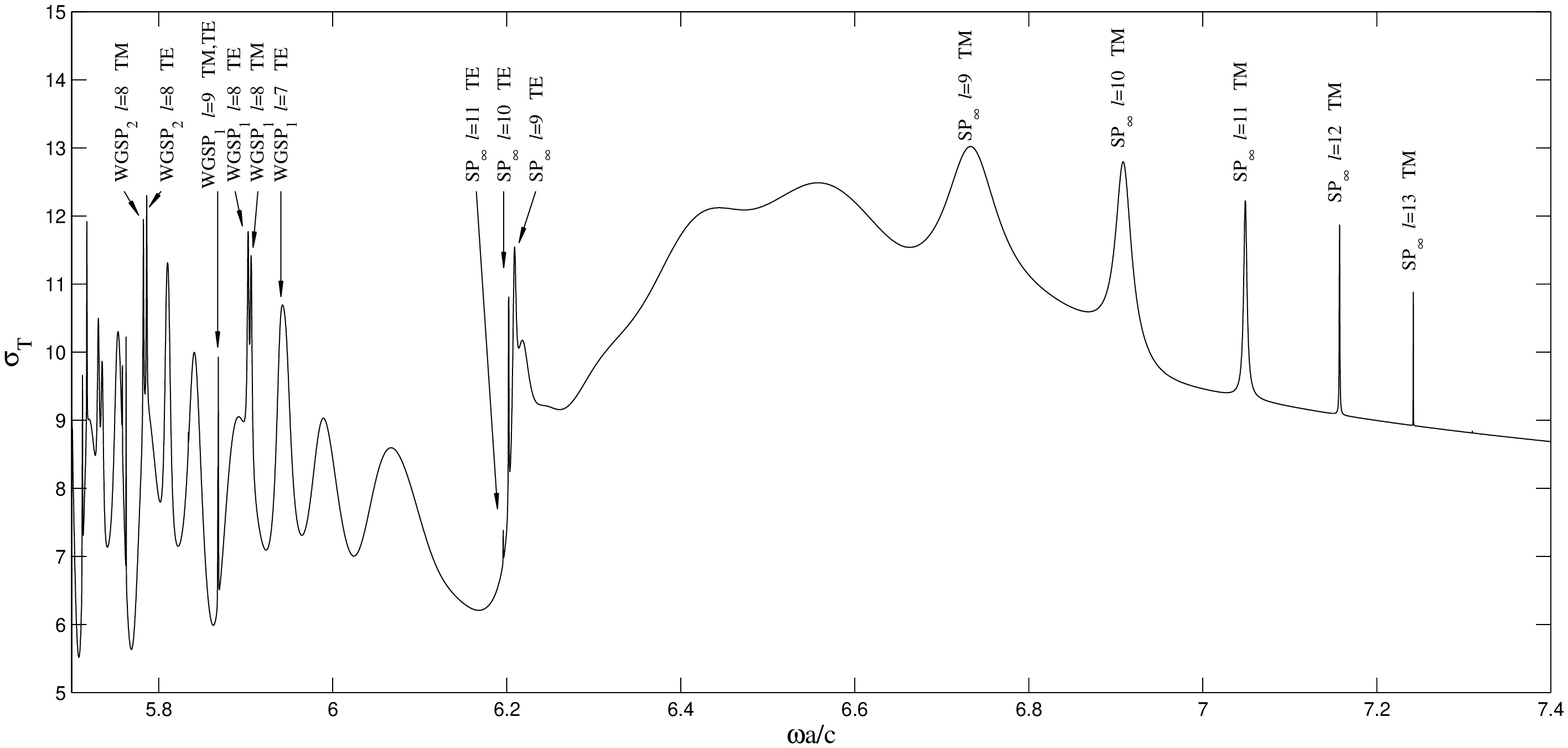}
\caption{\label{fig:ZoomTMTE} Zoom-in on the total cross section
$\sigma_T$.}
\end{figure*}

In Figs.~\ref{fig:RTSPiTM}-\ref{fig:RTWGSPTE}, we have displayed the
Regge trajectories of the first SP's for the TM and TE
polarizations. They have been obtained by solving numerically
Eq.~(\ref{RP}). We can observe some interesting features:

\qquad -- The dispersion curve for the surface wave
${\mathrm{SP}_\infty}$ of the TM polarization is a positive and
monotonically increasing function of $\omega$. As a consequence, the
associated group and phase velocities given by Eq.~(\ref{VpGg}) are
both positive and ${\mathrm{SP}_\infty}$ has an ordinary behavior.
It should be also noted that this SP exists in the frequency range
$\omega \in ]0, \omega_s[$ and therefore in the range $]\omega_0,
\omega_b[$ where the refraction index is negative but also outside
this range. For low values of $\omega$, its damping becomes very
large and thus this surface wave has a negligible role for these
frequencies in the scattering process and in the resonance
mechanism. Furthermore, as $\omega \to \omega_s$ the dispersion
curve increases indefinitely. This result will permit us to explain
the accumulation of resonances which converge to the limiting
frequency $\omega_s$ for the TM polarization.

\qquad -- The dispersion curve for the surface wave
${\mathrm{SP}_\infty}$ of the TE polarization is a positive and
monotonically decreasing function of $\omega$. As a consequence, the
associated phase velocity is positive while the group velocity is
negative [see Eq.~(\ref{VpGg})]. ${\mathrm{SP}_\infty}$ has a
``left-handed behavior". It should be also noted that this SP only
exists in the frequency range $]\omega_f, \omega_b[$ which is
included in the frequency range $]\omega_0, \omega_b[$ where the
refraction index is negative. Its damping is always weak and thus
this surface wave always plays a significant role in the scattering
process and in the resonance mechanism. Finally, as $\omega \to
\omega_f$ the dispersion curve increases indefinitely. This result
will permit us to explain the accumulation of resonances which
converge to the limiting frequency $\omega_f$ for the TE
polarization.

\qquad -- As far as the surface waves ${\mathrm{WGSP}_n}$ ($n \in
\mathbb{N}^*$) of the TM and TE polarizations are concerned, they
present, at first sight, a behavior which is rather independent of
the polarization. The real part of a given Regge pole
$\lambda_{\mathrm{WGSP}_n}$ vanishes for a frequency in the
 range $]\omega_0, \omega_b[$ and becomes negative. This
Regge pole then migrates to the third quadrant of the CAM plane and
becomes unphysical. So we can consider that the surface waves
${\mathrm{WGSP}_n}$ with $n \in \mathbb{N}^*$ only exist in a
subdomain of the frequency range $]\omega_0, \omega_b[$ where the
refraction index is negative. The dispersion relations of all these
surface waves are positive but monotonically decreasing functions.
Their group velocities are always negative while their phase
velocities are positive [see Eq.~(\ref{VpGg})]. All these SP's thus
have a left-handed behavior. Furthermore, because the dampings of
these surface waves are always weak, they all play a significant
role in the scattering process and in the resonance mechanism.
Finally, it should be noted that as $\omega \to \omega_0$, the
dispersion curves increase indefinitely. This result will permit us
to explain the accumulation of resonances which converges to the
limiting frequency $\omega_0$ for the TM and TE polarizations.

\subsection{Resonances}

In Figs.~\ref{fig:ResSPinfTM}-\ref{fig:ResWGSPTE} we present samples
of complex frequencies for the RSPM's associated with the surface
waves ${\mathrm{SP}_\infty}$ and ${\mathrm{WGSP}_n}$ with $n = 1$,
$2$ and $3$. They have been calculated from the semiclassical
formulas (\ref{sc1}) and (\ref{sc2}) by using the Regge trajectories
determined numerically by solving Eq.~(\ref{RP}) (see
Figs.~\ref{fig:RTSPiTM}-\ref{fig:RTWGSPTE}). A comparison between
the semiclassical spectra and the ``exact ones" (calculated by
solving numerically Eq.~\ref{det}) shows a very good agreement.
Moreover, we can also observe some interesting features ({\it
mutatis mutandis} they were already present for the left-handed
cylinder):

\qquad -- The resonance spectrum associated with the surface wave
${\mathrm{SP}_\infty}$ of the TM polarization (see
Fig.~\ref{fig:ResSPinfTM}) extends beyond the frequency range
$]\omega_0, \omega_b[$ where the sphere presents a left-handed
behavior because ${\mathrm{SP}_\infty}$ exists for $\omega \in ]0,
\omega_s[$. Furthermore, inserted into the semiclassical formulas
(\ref{sc1}) and (\ref{sc2}), the behavior of the Regge trajectory of
$\lambda_{\mathrm{SP}_\infty}$ near $\omega_s$ easily explains the
existence of the family of resonances close to the real axis of the
complex $\omega$ plane which converges for large $\ell$ to the
limiting frequency $\omega_s$.

\qquad -- The resonance spectrum associated with the surface wave
${\mathrm{SP}_\infty}$ of the TE polarization (see
Fig.~\ref{fig:ResSPinfTE}) fully lies inside the frequency range
$]\omega_0, \omega_b[$ where the sphere presents a left-handed
behavior because ${\mathrm{SP}_\infty}$ exists only in that range.
Furthermore, inserted into the semiclassical formulas (\ref{sc1})
and (\ref{sc2}), the behavior of the Regge trajectory of
$\lambda_{\mathrm{SP}_\infty}$ near $\omega_f$ explains the
existence of the family of resonances close to the real axis of the
complex $\omega$ plane which converges for large $\ell$ to the
limiting frequency $\omega_f$.

\qquad -- The resonance spectra associated with the surface waves
${\mathrm{WGSP}_n}$ ($n \in \mathbb{N}^*$) of the TM and TE
polarizations (see Figs.~\ref{fig:ResWGSPTM} and
\ref{fig:ResWGSPTE}) fully lie inside the frequency range
$]\omega_0, \omega_b[$ where the sphere presents a left-handed
behavior because all these surface waves exist only in that range.
Furthermore, inserted into the semiclassical formulas (\ref{sc1})
and (\ref{sc2}), the behavior of the Regge trajectory of a given
Regge pole $\lambda_{\mathrm{WGSP}_n}$ near $\omega_0$ explains the
existence of a corresponding family of resonances close to the real
axis of the complex $\omega$ plane which converges for large $\ell $
to the limiting frequency $\omega_0$.

\begin{figure}
\includegraphics[height=5.6cm,width=8cm]{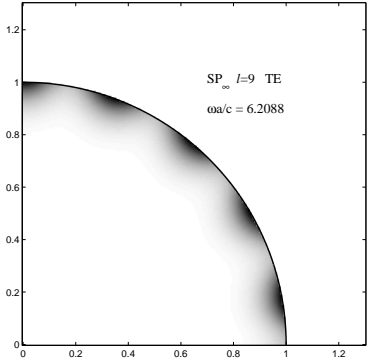}
\caption{\label{fig:DensEnergySPinf9} Electromagnetic energy density
for the resonant mode $\ell=9$ generated by ${\mathrm{SP}_\infty}$
(TE polarization).}
\end{figure}

\begin{figure}
\includegraphics[height=5.6cm,width=8cm]{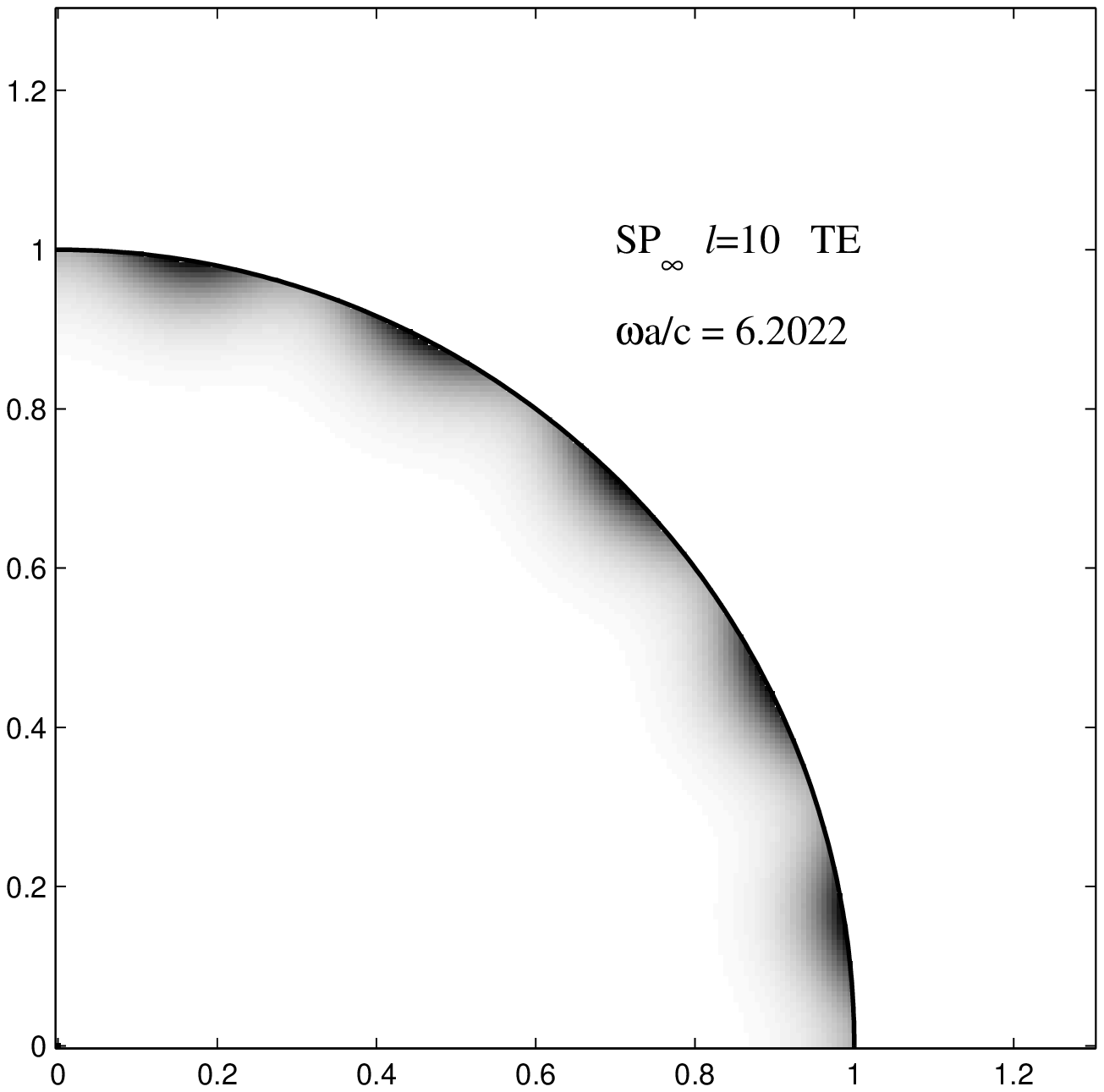}
\caption{\label{fig:DensEnergySPinf10} Electromagnetic energy
density for the resonant mode $\ell=10$ generated by
${\mathrm{SP}_\infty}$ (TE polarization).}
\end{figure}

\begin{figure}
\includegraphics[height=5.6cm,width=8cm]{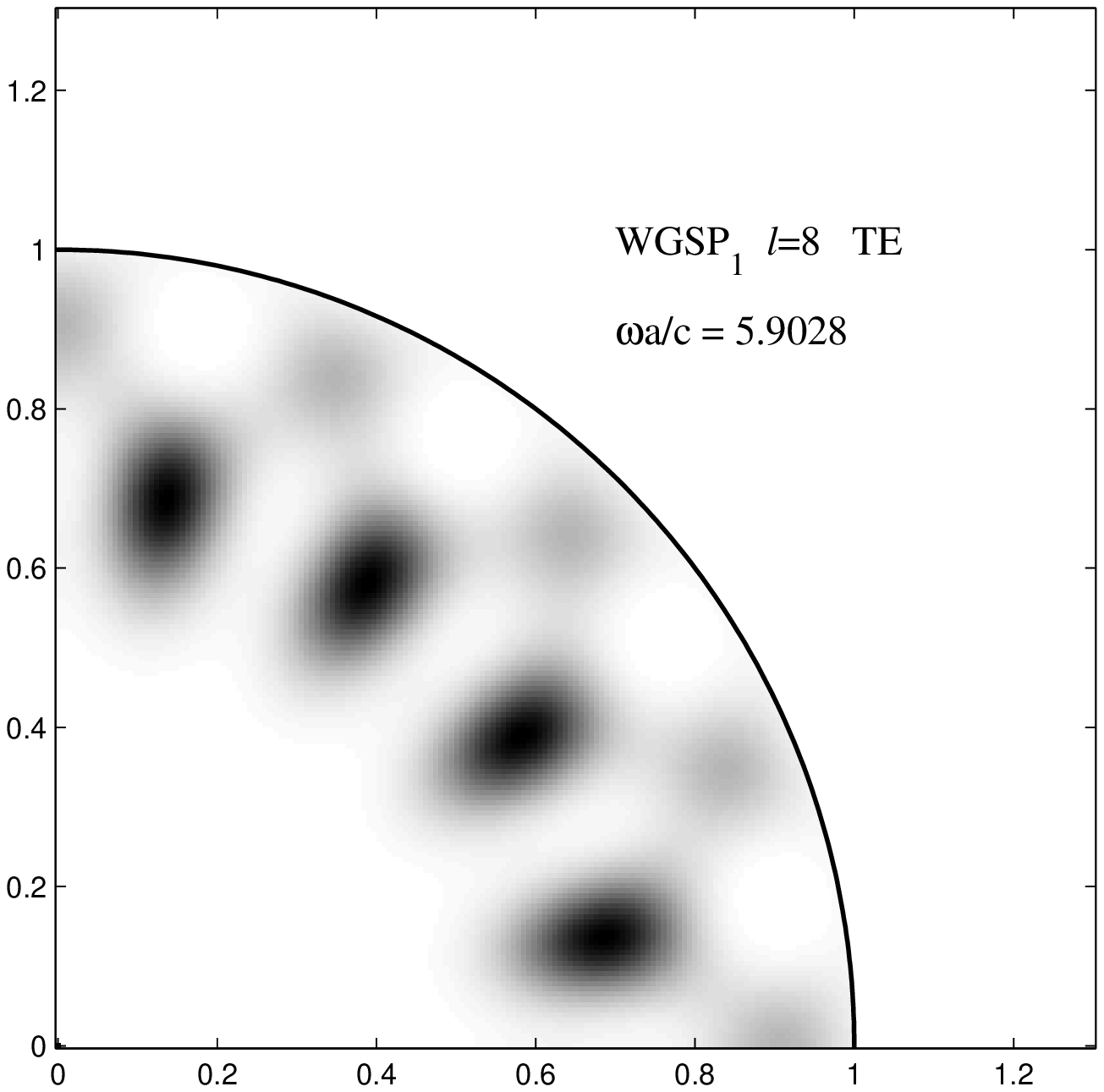}
\caption{\label{fig:DensEnergyWGSP8} Electromagnetic energy density
for the resonant mode $\ell=8$ generated by ${\mathrm{WGSP}_1}$ (TE
polarization).}
\end{figure}

\begin{figure}
\includegraphics[height=5.6cm,width=8cm]{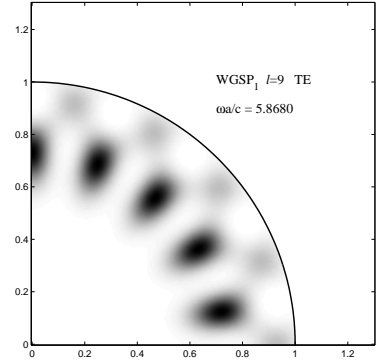}
\caption{\label{fig:DensEnergyWGSP9} Electromagnetic energy density
for the resonant mode $\ell=9$ generated by ${\mathrm{WGSP}_1}$ (TE
polarization).}
\end{figure}

In conclusion, we have established a connection between the complex
frequencies of the long-lived resonant modes (or RSPM's) of the
left-handed sphere and the SP's noted ${\mathrm{SP}_\infty}$ and
${\mathrm{WGSP}_n}$ with $n \in \mathbb{N}^*$ which are supported by
its surface. In other words, in spite of the great confusion which
seems to prevail in the resonance spectrum of the left-handed sphere
(see Sec. II), we have been able to fully classify and physically
interpret the resonances thanks to CAM techniques. We now invite the
reader to look at Fig.~\ref{fig:ZoomTMTE} where we have zoomed in on
Fig.~\ref{fig:general_TETM}. On the total cross sections $\sigma
_T(\omega)$ we have identified the peaks corresponding to resonances
and, for each one, we have specified the SP which has generated it
as well as its polarization and the associated ``quantum number"
$\ell$. This has been achieved by using the results displayed on
Figs.~\ref{fig:ResSPinfTM}-\ref{fig:ResWGSPTE}. It is also possible
to identify the ``quantum numbers" of the different resonant modes
by plotting the associated distribution of electromagnetic energy
density. Indeed, the number of local maxima is twice the quantum
number $\ell$. These distributions can been obtained from the theory
developed in Refs.~\onlinecite{Ruppin_Energy1998,Ruppin_Energy2002}.
For example, in
Figs.~\ref{fig:DensEnergySPinf9}-\ref{fig:DensEnergyWGSP9} we have
displayed, for the TE polarization, the energy distributions for the
resonant modes $\ell=9$ and $\ell=10$ generated by
${\mathrm{SP}_\infty}$ and for the resonant modes $\ell=8$ and
$\ell=9$ generated by ${\mathrm{WGSP}_1}$. For obvious symmetry
reasons, we have only considered one quarter of the equatorial
section of the sphere.



\section{Asymptotics for
surface polaritons and physical description}

In order to obtain a deeper physical understanding of the SP's
orbiting around the left-handed sphere and to justify the
terminology previously used (i.e., the notations
${\mathrm{SP}_\infty}$ and ${\mathrm{WGSP}_n}$ for the SP's), we
must ``analytically" solve Eq.~(\ref{RP}) for $\lambda =
\lambda_\mathrm{SP}$ or equivalently
\begin{equation} \label{RPSP1_TM}
 \sqrt{\frac{\epsilon(\omega)}{\mu(\omega)}} \frac{\zeta
 _{\lambda_\mathrm{SP}-1/2
}^{(1)^{\prime }}\left(  a\omega/c\right) }{\zeta
_{\lambda_\mathrm{SP}-1/2 }^{(1)} \left(
a\omega/c\right)}=\frac{\psi _{\lambda_\mathrm{SP}-1/2 }^{\prime
}\left[n(\omega) a\omega/c\right]}{\psi _{\lambda_\mathrm{SP}-1/2
}\left[n(\omega) a\omega/c\right]}
\end{equation}
for the TM polarization and
\begin{equation} \label{RPSP1_TE}
 \sqrt{\frac{\mu(\omega)}{\epsilon(\omega)}} \frac{\zeta _{\lambda_\mathrm{SP}-1/2
}^{(1)^{\prime }}\left(  a\omega/c\right) }{\zeta
_{\lambda_\mathrm{SP}-1/2 }^{(1)} \left(
a\omega/c\right)}=\frac{\psi _{\lambda_\mathrm{SP}-1/2 }^{\prime
}\left[n(\omega) a\omega/c\right]}{\psi _{\lambda_\mathrm{SP}-1/2
}\left[n(\omega) a\omega/c\right]}
\end{equation}
for the TE polarization. In other words, we seek to obtain explicit
formulas for the Regge poles of the problem. We have previously done
it for metallic and semiconducting objects\cite{ADFG_1,ADFG_3} as
well as for left-handed cylinders\cite{ADFG_2} and we shall here
extend such an analysis to left-handed spheres. We were recently
aware of a beautiful article by Berry\cite{BerryMV1975} where this
problem has also been considered for a curved dielectric interface
with an electric permittivity negative and independent of $\omega$.

By replacing Ricatti-Bessel functions by spherical Bessel functions
and then using their relations with the ordinary Bessel functions
(see Ref.~\onlinecite{AS65})
\begin{equation} \label{SpB-B}
j_\lambda (z)=\sqrt{\frac{\pi}{2z}}J_{\lambda+1/2}(z) \quad
\mathrm{and} \quad h^{(1)}_\lambda
(z)=\sqrt{\frac{\pi}{2z}}H^{(1)}_{\lambda+1/2}(z),
\end{equation}
as well as $J_\lambda (-z)=e^{i\pi \lambda }J_\lambda (z) $ and $
J'_\lambda (-z)=-e^{i\pi \lambda }J'_\lambda (z) $ (see
Ref.~\onlinecite{AS65}) in order to take into account that
$n(\omega)<0$ in the most interesting frequency range, it is easy to
prove that Eqs.~(\ref{RPSP1_TM}) and (\ref{RPSP1_TE}) respectively
reduce to
\begin{eqnarray} \label{RPSP2_TM}
& & -\frac{J'_{\lambda_\mathrm{SP}} [|n(\omega)| \, \omega
a/c]}{J_{\lambda_\mathrm{SP}}[|n(\omega)| \, \omega a/c]}+\frac{1}{2
n(\omega)} \left( \frac{c}{\omega a}\right) \nonumber
\\
& & \qquad =\sqrt{\frac{\epsilon (\omega)}{\mu (\omega)}}
\frac{H_{\lambda_\mathrm{SP}}^{(1)'} (\omega
a/c)}{H_{\lambda_\mathrm{SP}}^{(1)} (\omega a/c)}+\frac{1}{2}
\sqrt{\frac{\epsilon (\omega)}{\mu (\omega)}}\left( \frac{c}{\omega
a}\right)
\end{eqnarray}
for the TM polarization and
\begin{eqnarray} \label{RPSP2_TE}
& & -\frac{J'_{\lambda_\mathrm{SP}} [|n(\omega)| \, \omega
a/c]}{J_{\lambda_\mathrm{SP}}[|n(\omega)| \, \omega a/c]}+\frac{1}{2
n(\omega)} \left( \frac{c}{\omega a}\right) \nonumber
\\
& & \qquad =\sqrt{\frac{\mu (\omega)}{\epsilon (\omega)}}
\frac{H_{\lambda_\mathrm{SP}}^{(1)'} (\omega
a/c)}{H_{\lambda_\mathrm{SP}}^{(1)} (\omega a/c)}+\frac{1}{2}
\sqrt{\frac{\mu (\omega)}{\epsilon (\omega)}}\left( \frac{c}{\omega
a}\right)
\end{eqnarray}
for the TE polarization. These two last equations must be compared
respectively with Eqs.~(37) and (38) of Ref.~\onlinecite{ADFG_2}
which provide SP Regge poles for the left-handed cylinder. The first
term on the left-hand side and on the right-hand side of
Eqs.~(\ref{RPSP2_TM}) and (\ref{RPSP2_TE}) are exactly those
appearing in Eqs.~(37) and (38) of Ref.~\onlinecite{ADFG_2}. The
remaining terms are curvature corrections due to the change of
dimension. Equations (\ref{RPSP2_TM}) and (\ref{RPSP2_TE}) can be
solved following the method used in order to solve Eqs.~(37) and
(38) of Ref.~\onlinecite{ADFG_2}, i.e., by using asymptotic
analysis. As we have previously noted in Ref.~\onlinecite{ADFG_2},
the choice of the asymptotic expansions for the Bessel functions
strongly depends on the relative positions of the arguments
$|n(\omega)|\omega a/c$ and $\omega a/c$ with respect to the complex
order $\lambda_\mathrm{SP}$. In order to simplify the discussion, we
choose to describe theoretically SP's in the frequency ranges where
they generate the RSPM's with the longest lifetime (such modes are
the most important from the physical point of view). In other words,
we shall seek $\lambda_{\mathrm{SP}_\infty}(\omega)$ for the TM
polarization with $\omega$ in the neighborhood of $\omega_s$,
$\lambda_{\mathrm{SP}_\infty}(\omega)$ for the TE polarization with
$\omega$ in the neighborhood of $\omega_f$ and
$\lambda_{\mathrm{WGSP}_n}(\omega)$ for both polarizations with
$\omega$ in the neighborhood of $\omega_0$ . In fact, in spite of
these restrictions, we shall obtain asymptotic results valid in
large frequency ranges.

\subsection{Asymptotics for surface polaritons of ${\mathrm{SP}_\infty}$-type and physical
description}

 Let us first consider
the Regge pole associated with ${\mathrm{SP}_\infty}$ for the TM
polarization. We assume $\omega$ in the neighborhood of $\omega_s$
[but also that $\omega < \omega_b$ so that $n(\omega) <0$] and then
we can also assume that $\mathrm{Re} \,
\lambda_{\mathrm{SP}_\infty}(\omega) > \omega a/c
> |n(\omega)|\omega a/c$ and formally that $\mathrm{Re} \,
\lambda_{\mathrm{SP}_\infty}(\omega) \gg 1$ and $\mathrm{Re} \,
\lambda_{\mathrm{SP}_\infty}(\omega) \gg \mathrm{Im} \,
\lambda_{\mathrm{SP}_\infty}(\omega)$. As a consequence, we can use
the Debye asymptotic expansions for $J_\lambda(z)$ and
$H_\lambda^{(1)}(z)$ valid for large orders (see our previous
paper\cite{ADFG_2} for details as well as Appendix A of
Ref.~\onlinecite{Nuss65} or Ref.~\onlinecite{WatsonBessel}). We can
then write
\begin{equation} \label{lhsRPSP2_TM}
\frac{J'_{\lambda_{\mathrm{SP}_\infty}}[ |n(\omega)| \omega a/c
]}{J_{\lambda_{\mathrm{SP}_\infty}}[ |n(\omega)| \omega a/c ]} \sim
\frac{\left[ \lambda_{\mathrm{SP}_\infty}^2 - (|n(\omega)|\omega
a/c)^2 \right]^{1/2}}{|n(\omega)|(\omega a/c)}
\end{equation}
and
\begin{eqnarray} \label{rhsRPSP2_TMcoorStokes}
&  & \frac{H_{\lambda_{\mathrm{SP}_\infty}}^{(1)'}( \omega a/c
)}{H_{\lambda_{\mathrm{SP}_\infty}}^{(1)}( \omega a/c )} \sim -
\frac{\left[ \lambda_{\mathrm{SP}_\infty}^2 -
(\omega a/c)^2 \right]^{1/2}}{(\omega a/c)}  \nonumber \\
& &  \qquad  \times \left( 1- i  e^{2
\alpha(\lambda_{\mathrm{SP}_\infty}, \omega a/c)} \right)
\end{eqnarray}
where
\begin{equation}
 \alpha(\lambda ,z)   =  (\lambda^2 - z^2)^{1/2} -\lambda \, \ln
\left( \frac{\lambda + (\lambda^2 - z^2)^{1/2}}{z} \right).
\label{AsympDebyeIc}
\end{equation}
In Eq.~(\ref{rhsRPSP2_TMcoorStokes}), we have taken into account an
exponentially small contribution (the term $\exp[2
\alpha(\lambda_{\mathrm{SP}_\infty}, \omega a/c)]$) which lies
beyond all orders in perturbation theory. This term can be captured
by carefully taking into account Stokes phenomenon and is necessary
to extract the asymptotic expression of the imaginary part of
$\lambda_\mathrm{SP}$. [In Eq.~(\ref{rhsRPSP2_TMcoorStokes}) we have
given to the Stokes multiplier function the value $1/2$.] For more
precision, we refer to our previous paper\cite{ADFG_2} as well as to
Refs.~\onlinecite{Berry89,Dingle73,SegurTL91,BerryHowls90}. Now, by
inserting  (\ref{lhsRPSP2_TM}) and (\ref{rhsRPSP2_TMcoorStokes})
into Eq.~(\ref{RPSP2_TM}), we obtain an equation which can be
``easily" solved. We first neglect the exponentially small term and
this equation reduces to a fourth-order polynomial equation. We only
retain the physical solution which correctly modifies the formula
(47a) obtained in Ref.~\onlinecite{ADFG_2} for the corresponding SP
of the left-handed cylinder. It provides the real part of
$\lambda_{\mathrm{SP}_\infty}(\omega)$. We then take into account
the exponentially small term and we obtain perturbatively the
imaginary part of $\lambda_{\mathrm{SP}_\infty}(\omega)$. We finally
have

{\small \begin{widetext}
\begin{subequations}
\begin{eqnarray} \label{RPSPiTMas}
&  & \mathrm{Re} \,  \lambda_{\mathrm{SP}_\infty} (\omega ) \sim
\left( \frac{\omega a}{c}  \right) \times \nonumber \\
& &  \sqrt{ \frac{|\epsilon(\omega )| [|\epsilon(\omega )| +\mu
(\omega)]}{\epsilon ^2 (\omega )-1} + \frac{\epsilon ^2 (\omega
)+1}{4[1-|\epsilon(\omega )|]^2} \left(\frac{c}{\omega a}\right)^2
+\sqrt{ \frac{\epsilon ^2 (\omega )[|\epsilon(\omega )|\mu
(\omega)+1]}{[\epsilon ^2 (\omega )-1][1-|\epsilon(\omega
)|]^2}\left(\frac{c}{\omega a}\right)^2+ \frac{\epsilon ^2 (\omega
)}{4[1-|\epsilon(\omega )|]^4}\left(\frac{c}{\omega a}\right)^4 }}
 \nonumber \\
& &
\label{reRPSPiTMas}  \\
&  & \mathrm{Im} \, \lambda_{\mathrm{SP}_\infty}(\omega) \sim
\left(\frac{\epsilon ^2 (\omega ) }{ \epsilon ^2 (\omega )-1}
\right)  \times \frac{ \left(\mathrm{Re} \,
\lambda_{\mathrm{SP}_\infty} (\omega )\right)^2-\left( {\omega a/c}
\right)^2 }{\mathrm{Re} \, \lambda_{\mathrm{SP}_\infty} (\omega )}
\times \exp[2 \alpha(\mathrm{Re} \, \lambda_{\mathrm{SP}_\infty}
(\omega ) ,\omega a/c)]. \label{imRPSPiTMas}
\end{eqnarray}
\end{subequations}
\end{widetext}
}

Let us now consider the Regge pole associated with
${\mathrm{SP}_\infty}$ of the TE polarization. In order to describe
it, we must solve Eq.~(\ref{RPSP2_TE}) which only differs from
Eq.~(\ref{RPSP2_TM}) by the exchange of $\epsilon(\omega)$ and
$\mu(\omega)$. Here, we assume $\omega$ in the neighborhood of
$\omega_f$ and then we can also assume that $\mathrm{Re} \,
\lambda_{\mathrm{SP}_\infty}(\omega)
> |n(\omega)|\omega a/c > \omega a/c$ and formally that
$\mathrm{Re} \, \lambda_{\mathrm{SP}_\infty}(\omega) \gg 1$ and
$\mathrm{Re} \, \lambda_{\mathrm{SP}_\infty}(\omega) \gg \mathrm{Im}
\, \lambda_{\mathrm{SP}_\infty}(\omega)$. As a consequence, we can
use again the Debye asymptotic expansions for $J_\lambda(z)$ and
$H_\lambda^{(1)}(z)$ valid for large orders and the resolution of
Eq.~(\ref{RPSP2_TE}) can be modeled on that of Eq.~(\ref{RPSP2_TM}).
Formulas (\ref{lhsRPSP2_TM}) and (\ref{rhsRPSP2_TMcoorStokes})
remain valid and by inserting them into Eq.~(\ref{RPSP2_TE}), we
obtain

{\small \begin{widetext}
\begin{subequations}
\begin{eqnarray} \label{RPSPiTEas}
&  & \mathrm{Re} \,  \lambda_{\mathrm{SP}_\infty} (\omega ) \sim
\left( \frac{\omega a}{c}  \right) \times \nonumber \\
& &  \sqrt{ \frac{|\mu(\omega )| [|\mu(\omega )| +\epsilon
(\omega)]}{\mu ^2 (\omega )-1} + \frac{\mu ^2 (\omega
)+1}{4[1-|\mu(\omega )|]^2} \left(\frac{c}{\omega a}\right)^2
-\sqrt{ \frac{\mu ^2 (\omega )[|\mu(\omega )|\epsilon
(\omega)+1]}{[\mu ^2 (\omega )-1][1-|\mu(\omega
)|]^2}\left(\frac{c}{\omega a}\right)^2+ \frac{\mu ^2 (\omega
)}{4[1-|\mu(\omega )|]^4}\left(\frac{c}{\omega a}\right)^4 }}
 \nonumber \\
& &
\label{reRPSPiTEas}  \\
&  & \mathrm{Im} \, \lambda_{\mathrm{SP}_\infty}(\omega) \sim
\left(\frac{\mu ^2 (\omega ) }{ \mu ^2 (\omega )-1} \right) \times
\frac{ \left(\mathrm{Re} \, \lambda_{\mathrm{SP}_\infty} (\omega
)\right)^2-\left( {\omega a/c} \right)^2 }{\mathrm{Re} \,
\lambda_{\mathrm{SP}_\infty} (\omega )} \times \exp[2
\alpha(\mathrm{Re} \, \lambda_{\mathrm{SP}_\infty} (\omega ) ,\omega
a/c)]. \label{imRPSPiTEas}
\end{eqnarray}
\end{subequations}
\end{widetext}
}

\begin{figure}
\includegraphics[height=8cm,width=8.6cm]{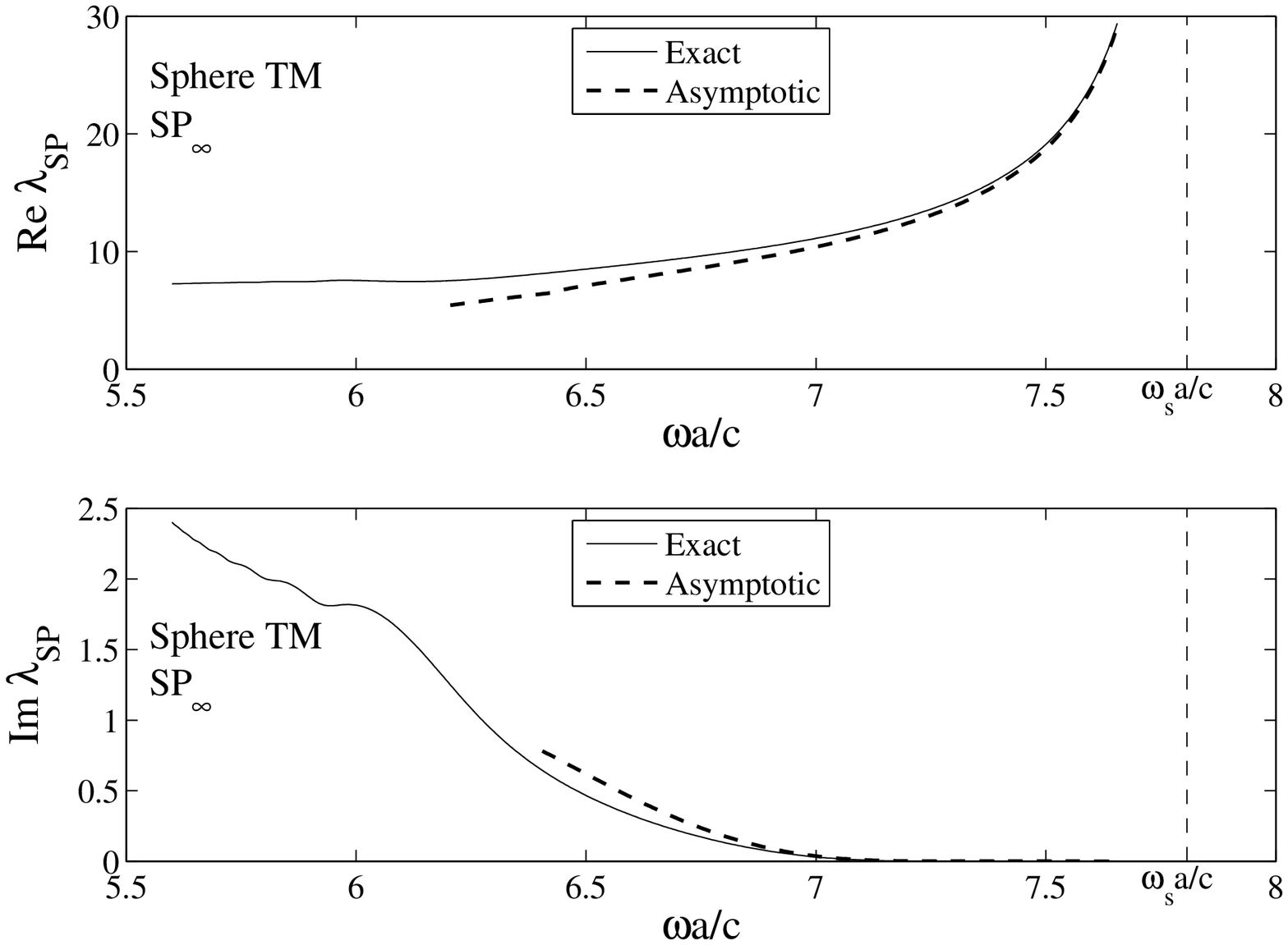}
\caption{\label{fig:RPSinfasTM} Regge trajectory for the Regge pole
associated with ${\mathrm{SP}_\infty}$ (TM polarization). Comparison
between exact and asymptotic theories.}
\end{figure}
\begin{figure}
\includegraphics[height=8cm,width=8.6cm]{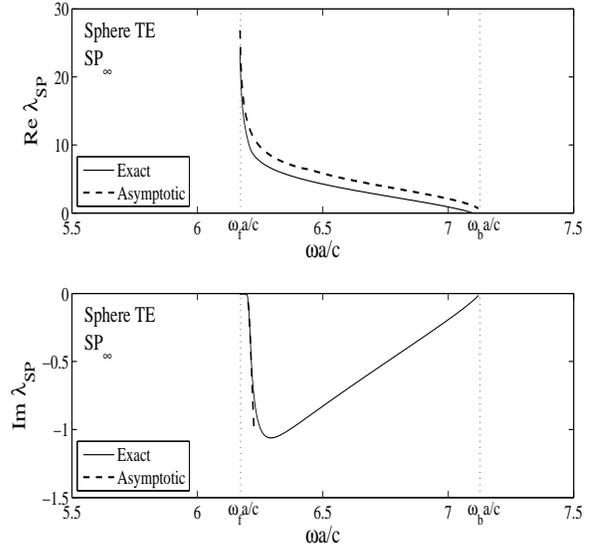}
\caption{\label{fig:RPSinfasTE} Regge trajectory for the Regge pole
associated with ${\mathrm{SP}_\infty}$ (TE polarization). Comparison
between exact and asymptotic theories.}
\end{figure}

Equations (\ref{reRPSPiTMas})-(\ref{imRPSPiTMas}) and
(\ref{reRPSPiTEas})-(\ref{imRPSPiTEas}) provide respectively
analytic expressions for the dispersion relation and the damping of
the surface polaritons ${\mathrm{SP}_\infty}$ of the TM and TE
polarizations. The following important physical features must be
noted:

\qquad -- These two SP's present exponentially small attenuations.

\qquad -- Their wave numbers $k_{\mathrm{SP}_\infty} (\omega)$ can
be obtained from (\ref{reRPSPiTMas}) and (\ref{WNSP}) for the TM
polarization and from (\ref{reRPSPiTEas}) and (\ref{WNSP}) for the
TE polarization and they could permit us to derive analytically
their phase velocity $v_p= \omega / k_{\mathrm{SP}_\infty}(\omega )$
as well as their group velocity $v_g= d\omega /
dk_{\mathrm{SP}_\infty}(\omega )$.

\qquad -- For $a \to \infty$ -- i.e., in the flat interface limit --
the wave number $k_{\mathrm{SP}_\infty} (\omega)$ for the TM
polarization reduces to
\begin{equation}\label{WNSPinfTM}
k_{\mathrm{SP}_\infty} (\omega) \sim \left( \frac{\omega}{c} \right)
\sqrt{ \frac{|\epsilon(\omega )| [|\epsilon(\omega )| +\mu
(\omega)]}{\epsilon ^2 (\omega )-1}}.
\end{equation}
This expression is the usual dispersion relation found in
Refs.~\onlinecite{RuppinPLA00,Darmanyanetal03,ShadrivovEtAl04} for
the $p$-polarized SP - i.e., the SP for which the magnetic field
$\mathbf{H}$ is normal to the incidence plane - supported by the
flat interface. For the TM polarization, ${\mathrm{SP}_\infty}$ is
therefore the counterpart of the $p$-polarized SP supported by the
flat interface.

\qquad -- For $a \to \infty$ -- i.e., in the flat interface limit --
the wave number $k_{\mathrm{SP}_\infty} (\omega)$ for the TE
polarization reduces to
\begin{equation}\label{WNSPinfTE}
k_{\mathrm{SP}_\infty} (\omega) \sim \left( \frac{\omega}{c} \right)
\sqrt{ \frac{|\mu(\omega )| [|\mu(\omega )| +\epsilon (\omega)]}{\mu
^2 (\omega )-1}}.
\end{equation}
This expression is the usual dispersion relation found in
Refs.~\onlinecite{RuppinPLA00,Darmanyanetal03,ShadrivovEtAl04} for
the $s$-polarized SP - i.e., the SP for which the electric field
$\mathbf{E}$ is normal to the incidence plane - supported by the
flat interface. For the TE polarization, ${\mathrm{SP}_\infty}$ is
therefore the counterpart of the $s$-polarized SP supported by the
flat interface.

\qquad -- Furthermore, for $a \to \infty$, the imaginary parts
(\ref{imRPSPiTMas}) and (\ref{imRPSPiTEas}) of
$\lambda_{\mathrm{SP}_\infty}$ vanish. In the flat interface limit,
$\mathrm{SP}_\infty$ for both polarizations have no damping like the
SP's supported by the flat
interface\cite{RuppinPLA00,Darmanyanetal03,ShadrivovEtAl04}.

\qquad -- A connection with the results obtained in our previous
study concerning the left-handed cylinder\cite{ADFG_2} can be
established. By comparing (\ref{reRPSPiTMas}) with (47a) of
Ref.~\onlinecite{ADFG_2} and (\ref{reRPSPiTEas}) with (49a) of
Ref.~\onlinecite{ADFG_2} we can notice that ${\mathrm{SP}_\infty}$
of the sphere for the TM (TE) polarization is the counterpart of
${\mathrm{SP}_\infty}$ of the cylinder for the TE (TM) polarization.
In Eqs.~(\ref{reRPSPiTMas}) and (\ref{reRPSPiTEas}), the
supplementary terms are associated with the curvature corrections
appearing in Eqs.~(\ref{RPSP2_TM}) and (\ref{RPSP2_TE}). In certain
frequency ranges, these corrections are of the same magnitude than
the remaining terms. As a consequence, it is not possible to obtain
dispersion relations on the sphere perturbatively from those on the
cylinder.

\qquad -- The function $\mathrm{Re}\, \lambda_{\mathrm{SP}_\infty}
(\omega )$ given by (\ref{reRPSPiTMas}) has a pole when
$\epsilon(\omega)+1=0$ -- i.e., for $\omega = \omega_s$.
Furthermore, the imaginary part (\ref{imRPSPiTMas}) of
$\lambda_{\mathrm{SP}_\infty}$ vanishes for $\omega =\omega_s$.
These two results explain semiclassically, for the TM polarization,
the accumulation of resonances which converge to the limiting
frequency $\omega_s$.

\qquad -- The function $\mathrm{Re} \, \lambda_{\mathrm{SP}_\infty}
(\omega )$ given by (\ref{reRPSPiTEas}) has a pole when
$\mu(\omega)+1=0$ -- i.e., for $\omega = \omega_f$. Furthermore, the
imaginary part (\ref{imRPSPiTEas}) of $\lambda_{\mathrm{SP}_\infty}$
vanishes for $\omega =\omega_f$. These two results explain
semiclassically, for the TE polarization, the accumulation of
resonances which converge to the limiting frequency $\omega_f$.

\qquad -- We have numerically tested formulas (\ref{reRPSPiTMas})
and (\ref{imRPSPiTMas}) (see Fig.~\ref{fig:RPSinfasTM}) as well as
formulas (\ref{reRPSPiTEas}) and (\ref{imRPSPiTEas}) (see
Fig.~\ref{fig:RPSinfasTE}). They provide rather good approximations
i) for $\mathrm {Re} \, \lambda_{\mathrm{SP}_\infty}(\omega)$ in
large frequency ranges for both polarizations, ii) for $\mathrm {Im}
\, \lambda_{\mathrm{SP}_\infty}(\omega)$ in the neighborhood  of
$\omega_s$ for the TM polarization and iii) for $\mathrm {Im} \,
\lambda_{\mathrm{SP}_\infty}(\omega)$ in the neighborhood  of
$\omega_f$ for the TE polarization.

\subsection{Asymptotics for surface polaritons of ${\mathrm{WGSP}_n}$ type and physical
description}

Let us finally consider the Regge poles associated with the surface
waves ${\mathrm{WGSP}_n}$ for the TM and TE polarizations. We must
now solve Eqs.~(\ref{RPSP2_TM}) and (\ref{RPSP2_TE}) for
$\lambda_\mathrm{SP}=\lambda_{\mathrm{WGSP}_n}$ by assuming $\omega$
in the neighborhood of $\omega_0$. The configuration is identical to
that of Ref.~\onlinecite{ADFG_2} for the left-handed cylinder. We
can consider that $\mathrm{Re} \, \lambda_{\mathrm{WGSP}_n}(\omega)
> \omega a/c $ and formally that $\mathrm{Re} \,
\lambda_{\mathrm{WGSP}_n}(\omega) \gg 1$ and $\mathrm{Re} \,
\lambda_{\mathrm{WGSP}_n}(\omega) \gg \mathrm{Im} \,
\lambda_{\mathrm{WGSP}_n}(\omega)$. Furthermore,
$\lambda_{\mathrm{WGSP}_n}$ is in the immediate neighborhood of
$|n(\omega)|\omega a/c$ i.e., lies in the Airy circle centered on
$|n(\omega)|\omega a/c$. As a consequence, we can still replace
$H^{(1)}_\lambda (z)$ by its Debye asymptotic expansion valid for
large orders and therefore insert (\ref{rhsRPSP2_TMcoorStokes}) into
(\ref{RPSP2_TM}) and (\ref{RPSP2_TE}). As far as $J_{\lambda} (z)$
is concerned, we must now use its uniform asymptotic expansion (see
Appendix A of Ref.~\onlinecite{Nuss65} or
Ref.~\onlinecite{WatsonBessel})
\begin{equation}
J_{\lambda} (z) {\sim}  \left( 2/z \right)^{1/3} Ai \left[ \left(
2/z \right)^{1/3}(\lambda -z) \right]   \label{AiryBesselJ}
\end{equation}
where $Ai (z)$ denotes the Airy function\cite{AS65}. Thus, we must
insert
\begin{eqnarray} \label{lhsRPWGSP}
 &  & \frac{J_{\lambda_{\mathrm{WGSP}_n}}^{'}[ |n(\omega)| \omega a/c
]}{J_{\lambda_{\mathrm{WGSP}_n}}[ |n(\omega)| \omega a/c ]} \sim
\lbrace{2/[|n(\omega)| \omega a/c ]\rbrace}^{1/3}  \nonumber \\ &  &
 \times \frac{{Ai}^{'} \left[ \lbrace{2/[|n(\omega)| \omega a/c ]\rbrace}^{1/3}
 [\lambda_{\mathrm{WGSP}_n} - |n(\omega)| \omega a/c ]
\right]}{Ai \left[ \lbrace{2/[|n(\omega)| \omega a/c
]\rbrace}^{1/3}[\lambda_{\mathrm{WGSP}_n} - |n(\omega)| \omega a/c
] \right]}.  \nonumber  \\
\end{eqnarray}
into (\ref{RPSP2_TM}) and (\ref{RPSP2_TE}). We obtain two equations
which can be solved by using the method already considered in our
previous work\cite{ADFG_2} and which is an adaptation of the method
invented by Rayleigh a long time ago in order to describe
mathematically the whispering-gallery phenomenon in acoustics
\cite{Rayleigh1887, Rayleigh1910} (see also
Ref.~\onlinecite{StreiferKodis}). After having noticed that the
curvature corrections can be neglected in that configuration, we
recover the formulas obtained for the left-handed cylinder. We have
\begin{subequations}\label{RPWGS TM_as}
\begin{eqnarray}
&  & \mathrm{Re} \,  \lambda_{\mathrm{WGSP}_n} (\omega ) \sim |n
(\omega)|  \omega a/c + \left( \frac{|n (\omega)| \omega a /c}{2}
\right)^{1/3} x_n
\nonumber\\
& & \label{reRPWGSP_TM_as} \\
& &    \mathrm{Im} \, \lambda_{\mathrm{WGSP}_n}(\omega) \sim
\sqrt{\frac{\epsilon(\omega)}{\mu(\omega)}} \left( \frac{|n
(\omega)| ~\omega a
/c}{2} \right)^{2/3}  \nonumber \\
& & \qquad \qquad  \times  \frac{\left[ \left[\mathrm{Re} \,
\lambda_{\mathrm{WGSP}_n} (\omega )\right]^2-
\left( {\omega a/c} \right)^2 \right]^{1/2}}{\left({\omega a/c}\right) x_n} \nonumber \\
& & \qquad \qquad  \times \exp \left[ 2 \alpha(\mathrm{Re} \,
\lambda_{\mathrm{WGSP}_n}(\omega ), \omega a/c) \right]
\label{imRPWGSP_TM_as}
\end{eqnarray}
\end{subequations}
for the TM polarization and
\begin{subequations} \label{RPWGS TE_as}
\begin{eqnarray}
&  & \mathrm{Re} \,  \lambda_{\mathrm{WGSP}_n} (\omega ) \sim |n
(\omega)|  \omega a/c + \left( \frac{|n (\omega)| \omega a /c}{2}
\right)^{1/3} x_n
\nonumber\\
& & \label{reRPWGSP_TE_as} \\
 &  & \mathrm{Im} \,
\lambda_{\mathrm{WGSP}_n}(\omega) \sim
\sqrt{\frac{\mu(\omega)}{\epsilon(\omega)}} \left( \frac{|n
(\omega)| ~\omega a
/c}{2} \right)^{2/3}  \nonumber \\
& & \qquad \qquad  \times  \frac{\left[ \left[\mathrm{Re} \,
\lambda_{\mathrm{WGSP}_n} (\omega )\right]^2-
\left( {\omega a/c} \right)^2 \right]^{1/2}}{\left({\omega a/c}\right) x_n} \nonumber \\
& & \qquad \qquad \times \exp \left[ 2 \alpha(\mathrm{Re} \,
\lambda_{\mathrm{WGSP}_n}(\omega ), \omega a/c) \right]
\label{imRPWGSP_TE_as}
\end{eqnarray}
\end{subequations}
for the TE polarization. In
Eqs.~(\ref{reRPWGSP_TM_as})-(\ref{imRPWGSP_TE_as}), we have
introduced the zeros $x_n$ of the Airy function (the first three
ones are $x_1\approx -2.3381...$, $x_2\approx -4.0879...$ and
$x_3\approx -5.5205...$).

\begin{figure}
\includegraphics[height=8cm,width=8.6cm]{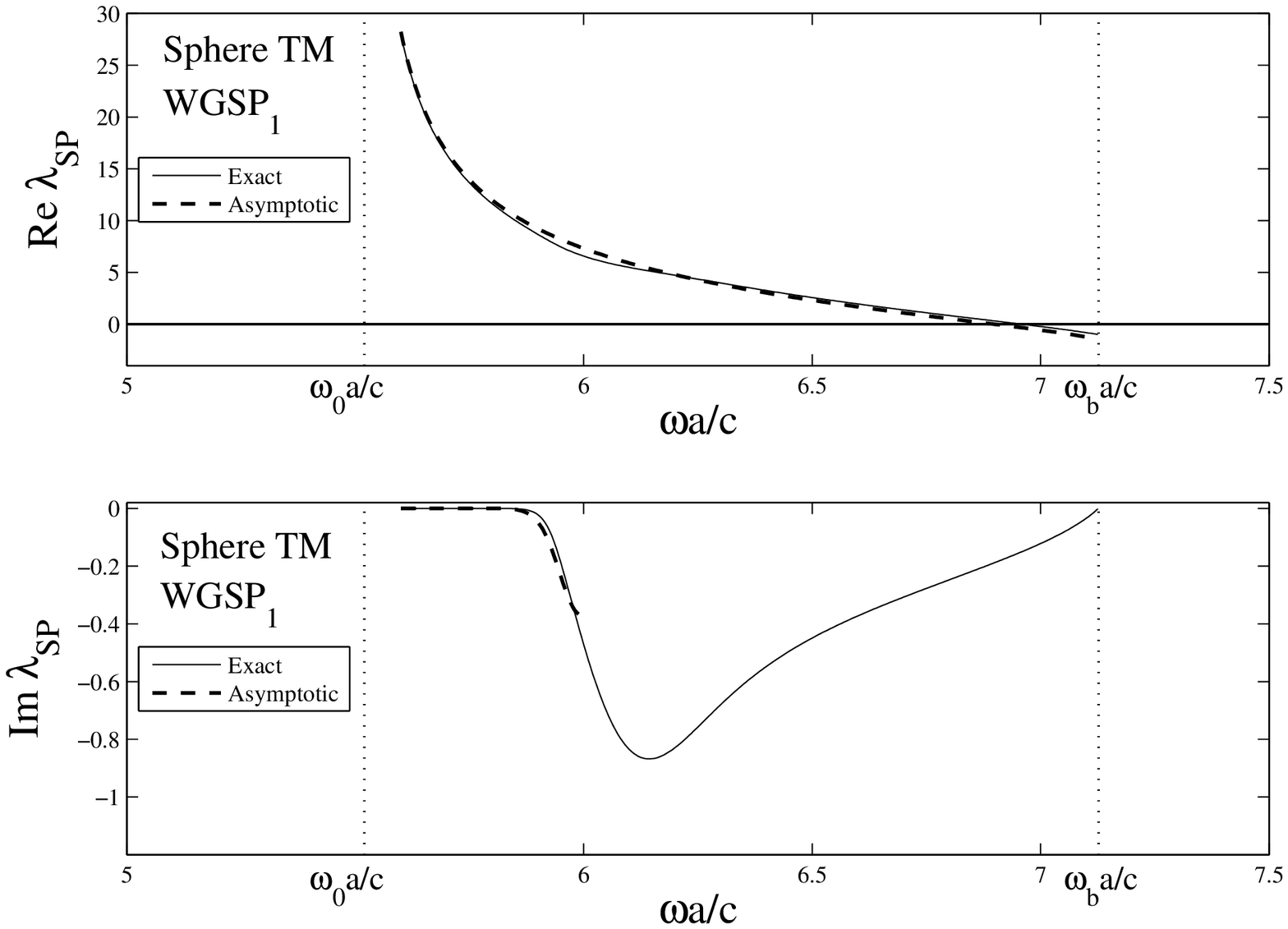}
\caption{\label{fig:RPWGSPasTM} Regge trajectory for the Regge pole
associated with ${\mathrm{WGSP}_1}$ (TM polarization). Comparison
between exact and asymptotic theories.}
\end{figure}
\begin{figure}
\includegraphics[height=8cm,width=8.6cm]{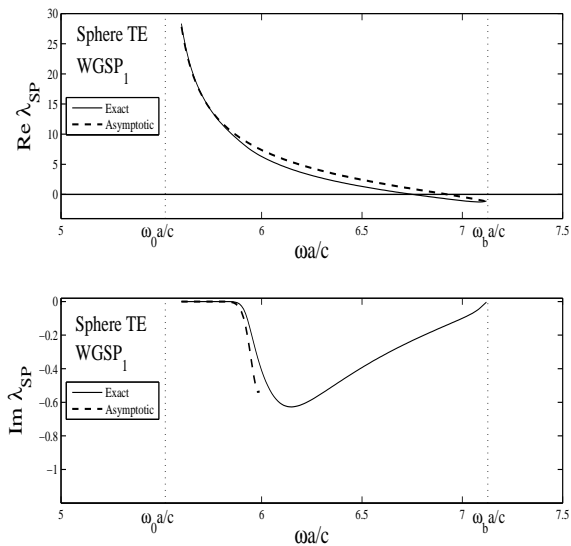}
\caption{\label{fig:RPWGSPasTE}Regge trajectory for the Regge pole
associated with ${\mathrm{WGSP}_1}$ (TE polarization). Comparison
between exact and asymptotic theories.}
\end{figure}

Equations (\ref{reRPWGSP_TM_as})-(\ref{imRPWGSP_TE_as}) provide
analytic expressions for the dispersion relation and the damping of
the surface polaritons ${\mathrm{WGSP}_n}$ for both polarizations.
These expressions are those already obtained for the left-handed
cylinder with the polarizations exchanged: (\ref{reRPWGSP_TM_as})
and (\ref{imRPWGSP_TM_as}) must be compared with (55a) and (55b) of
Ref.~\onlinecite{ADFG_2} while (\ref{reRPWGSP_TE_as}) and
(\ref{imRPWGSP_TE_as}) must be compared with (55a) and (55c) of
Ref.~\onlinecite{ADFG_2}. Therefore, {\it mutatis mutandis}, the
physical remarks and analysis of Ref.~\onlinecite{ADFG_2} concerning
the whispering-gallery surface waves propagating on the cylinder can
be taken again. We just note that:

\qquad -- These SP's have no counterparts in the plane interface
case.

\qquad -- They do not exist for curved metallic or semiconducting
interfaces (see Refs.~\onlinecite{ADFG_1,ADFG_3}).

\qquad -- The function $\mathrm{Re} \, \lambda_{\mathrm{WGSP}_n}
(\omega )$ given by (\ref{reRPWGSP_TM_as}) or (\ref{reRPWGSP_TE_as})
has a pole which is that of $n(\omega)$ and therefore which
corresponds to $\omega = \omega_0$. Furthermore, the imaginary parts
(\ref{imRPWGSP_TM_as}) and (\ref{imRPWGSP_TE_as}) of
$\lambda_{\mathrm{WGSP}_n}$ vanish for $\omega =\omega_0$. These
results justify all our previous remarks concerning the
accumulations of resonances which converge to the limiting frequency
$\omega_0$.

\qquad -- We have numerically tested formulas
(\ref{reRPWGSP_TM_as})-(\ref{imRPWGSP_TE_as}) (see
Figs.~\ref{fig:RPWGSPasTM} and \ref{fig:RPWGSPasTE} for
$\mathrm{WGSP}_1$). They provide very good approximations for
$\mathrm {Re} \, \lambda_{\mathrm{WGSP}_1}(\omega)$ in the full
frequency range $]\omega_0,\omega_b[$ where the sphere presents a
left-handed behavior.  They also provide very good approximations
for $\mathrm {Im} \, \lambda_{\mathrm{WGSP}_1}(\omega)$ in a rather
large frequency range above the limiting frequency $\omega_0$.

\section{Conclusion and perspectives}

In the present paper, we have considered the interaction of
electromagnetic waves with a sphere fabricated from a left-handed
material. We have mainly emphasized the resonant aspects of the
problem. We have shown that the long-lived resonant modes can be
classified into distinct families, each family being generated by
one SP and we have physically described all the SP's orbiting around
the sphere by providing, for each one, a numerical and a
semiclassical description of its dispersion relation and its
damping.

We have more particularly shown that the left-handed spherical
interface can support both TE and TM polarized SP's. For each
polarization, there exists a particular SP which corresponds, in the
large radius limit, to the SP which is supported by the plane
interface and which has been theoretically described in
Refs.~\onlinecite{RuppinPLA00,Darmanyanetal03,ShadrivovEtAl04}.
However, there also exists, for each polarization, an infinite
family of SP's of whispering gallery type and these have no
analogues in the plane interface case. They are only supported by
the curved left-handed interface (they do not exist for curved
metallic or semiconducting interfaces\cite{ADFG_1,ADFG_3}) and they
are analogs to those described for the cylindrical interface in
Ref.~\onlinecite{ADFG_2}.

One could believe at first sight that the existence of surface waves
of whispering gallery type (and of the associated resonant modes) in
the frequency region where both permittivity and permeability of the
sphere are negative is rather trivial because they are also present
for an ordinary dielectric sphere. So, one could also believe that
it is inappropriate to consider such surface waves as SP's. In fact,
that is not at all the case. Indeed, it is well known that the
excitation frequencies $\omega^{(0)}_{\ell n}$ of the
whispering-gallery resonant modes of an ordinary dielectric sphere
having a positive and frequency-independent refractive index $n$ are
obtained from the relation (see e.g.
Refs.~\onlinecite{LamLeungYoung1992,Schiller1993})
\begin{eqnarray} \label{res_wgm_ordin}
& & n ~ \omega a/c = \ell +1/2 - \left(\frac{\ell
+1/2}{2}\right)^{1/3} x_n \nonumber \\
& & \quad  + \hbox{corrections depending on the polarization}.
\end{eqnarray}
By replacing $n$ by $n(\omega)$, this relation permits us to
describe the scattering resonances of the left-handed sphere when
both permittivity and permeability are positive and therefore the
scattering resonances appearing on Fig.~\ref{fig:general_TETM} for
$\omega>\omega_p$. This result can be obtained by extending the
approach developed in
Refs.~\onlinecite{LamLeungYoung1992,Schiller1993}. On the other
hand, one can easily verify that this relation does not provide the
excitation frequencies of the whispering-gallery resonant modes when
the sphere presents the left-handed behavior (for
$\omega_0<\omega<\omega_b$ on Fig.~\ref{fig:general_TETM}). In this
case, it is necessary to use the theory which we developed in Sec.
IV. From (\ref{reRPWGSP_TM_as}) or (\ref{reRPWGSP_TE_as}) which have
been obtained by assuming $\epsilon(\omega)<0$ and $\mu(\omega)<0$
and from (\ref{sc1}) we then obtain
\begin{eqnarray} \label{res_wgm_lhm}
& & -n(\omega) \omega a/c = \ell +1/2 - \left(\frac{\ell
+1/2}{2}\right)^{1/3} x_n + \dots
\end{eqnarray}
Equations (\ref{res_wgm_ordin}) and (\ref{res_wgm_lhm}) seem similar
in form because they both describe whispering-gallery resonant modes
but, in fact, they are very different. They have distinct ranges of
validity and thus describe physical phenomena of different nature
(right-handed behavior in the first case and left-handed behavior as
well as accumulation of resonances in the second one). The existence
of the left-handed whispering-gallery resonant modes is related to
the conditions $\epsilon(\omega)<0$ and $\mu(\omega)<0$ and
therefore, more particularly (but not only), to the existence of
free electrons. As a consequence, we think that these modes can be
called resonant surface polariton modes and that the associated
surface waves described by (\ref{RPWGS TM_as}) and (\ref{RPWGS
TE_as}) are of SP type.

Because of the proliferation of SP's, left-handed spheres are
systems much richer than metallic or semiconducting
spheres\cite{ADFG_3}. They are therefore much more interesting as
artificial atoms (``plasmonic atoms")
\cite{Sakoda2001,ADFG_1,ADFG_3,GuzatovKlimov2007} and this could
have important consequences in terms of practical applications in
the fields of plasmonics and nanotechnologies. Here, we shall
briefly more particularly focus our discussion to the possible
applications to cavity quantum electrodynamics. It is well known
that high quality factor whispering-gallery modes in microdisks and
microspheres, because they are associated with a strong localization
of the electromagnetic field and a very long lifetime for the
photons, make these resonators interesting for practical
applications (see, for example, Ref.~\onlinecite{vonKlitzing01} and
references therein). They have been used, in particular, to produce
lasers and to enhance the spontaneous emission of quantum dots. Up
to now, these resonators have been fabricated from ordinary solid
dielectric materials. We think that the use of left-handed material
could constitute an interesting appealing alternative: indeed,
whispering-gallery modes supported by left-handed disks or spheres
seem  to present the interesting usual properties (strong
localization and very long lifetime) but also very unusual ones
linked with the left-handed behavior (negative group velocities and
positive phase velocities). However, at this stage of the
reflection, it is not possible for us to say if quantum
electrodynamics in ``left-handed resonators" could generate new
interesting physical phenomena and lead to advances in physics.

Finally, it should be noted that our approach as well as our main
results are not limited to the left-handed materials described by
the effective electric permittivity and the effective magnetic
permeability (\ref{PetP1}) and (\ref{PetP2}) but still remain valid
for more general left-handed materials (see the conclusion of
Ref.~\onlinecite{ADFG_2}).

\bibliography{SPsphereLHM}

\end{document}